\documentclass[letterpaper,11pt,leqno]{article}

\usepackage{endnotes}
\usepackage{amssymb}
\usepackage{amsfonts}
\usepackage{amsmath}
\usepackage{amsthm}
\usepackage{graphicx,color}

\usepackage{comment}
\usepackage{chicago}

\usepackage{array}

\usepackage{url}

\usepackage{setspace}
\def\R{\mathbb{R}}
\def\endproof{\hfill\diamondsuit}

\def\sF{{\mathcal F}}

\def\sA{{\mathcal A}}
\def\sL{{\mathcal L}}

\def\E{\mathbb{E}}

\def\sF{\mathcal{F}}
\def\P{\mathbb{P}}

\def\Q{\mathbb{Q}}

\def\sE{{\mathcal E}}

\numberwithin{equation}{section}
\theoremstyle{plain}                % title and  number in bold, text italic
\newtheorem{theorem}{Theorem}[section]
\newtheorem{lemma}[theorem]{Lemma}

\theoremstyle{definition}           % title and number in bold, text normal
\newtheorem{definition}[theorem]{Definition}

\newtheorem{conjecture}[theorem]{Conjecture}
\newtheorem{assumption}[theorem]{Assumption}
\theoremstyle{remark}               % title and number in italic, text normal

\newcommand{\papertitle}{Incomplete Continuous-time Securities Markets \\with Stochastic Income Volatility}

\begin{document}
\begin{titlepage}\renewcommand{\thefootnote}{\alph{footnote}}
\pagestyle{empty}
\begin{center}
\LARGE{\bf \papertitle}\footnote{We wish to thank Torben G. Andersen, Jerome Detemple, Darrell Duffie, Semyon Malamud, Claus Munk, Mark Schroder, Costis Skiadas, Steve Shreve, and Gordan {\v Z}itkovi\'c as well as the participants at the following conferences: the opening meeting for the math-finance semester at the Fields institute (2010), the SIAM math-finance meeting in San Francisco (2010), the Oberwolfach meeting (2011), and the SAFI meeting in Michigan (2011) for constructive comments.}
\ \\ \ \\
\end{center}
\begin{center}

{\large \bf Peter O. Christensen}\\ Department of Economics and Business, \\
 Aarhus University,\\ DK-8000 Aarhus C, Denmark  \\ email: {\tt
    pochristensen@econ.au.dk}

%\vspace*{2cm}
\ \\
{\large \bf Kasper Larsen}\\ Department of Mathematical Sciences, \\
  Carnegie Mellon University,\\ Pittsburgh, PA 15213, USA \\ email: {\tt
    kasperl@andrew.cmu.edu}

\end{center}
\begin{center}
\ \\ \
{\normalsize \today }
\end{center}
\ \\
%\vspace*{0.5cm}
%\newpage
\begin{verse}
{\sc Abstract}: In an incomplete continuous-time securities market with uncertainty generated by Brownian motions, we derive closed-form solutions for the equilibrium interest rate and market price of risk processes. The economy has a finite number of heterogeneous exponential utility investors, who receive partially unspanned income and can trade continuously on a finite time-interval in a money market account and a single risky security. Besides establishing the existence of an equilibrium, our main result shows that if the investors' unspanned income has stochastic countercyclical volatility, the resulting equilibrium can display both lower interest rates and higher risk premia compared to the Pareto efficient equilibrium in an otherwise identical complete market.
%Consequently, our model can simultaneously help explaining the risk-free rate and equity premium puzzles.
%We examine a class of Brownian based models which produce tractable incomplete equilibria. The models are based on finitely many investors with heterogeneous exponential utilities over intermediate consumption who receive partially unspanned income. The investors can trade continuously on a finite time interval in a money market account as well as a risky security. Besides establishing the existence of an equilibrium, our main result shows that the resulting equilibrium can display a lower risk-free rate and a higher risk premium relative to the usual Pareto efficient equilibrium in complete markets. Consequently, our model can simultaneously help explaining the risk-free rate and equity premium puzzles.
\end{verse}
%\vspace{0.5cm}
\begin{verse}
\end{verse}
%\begin{verse}
%{\sc AMS subject classifications}: 93E20\\
%\vspace*{2ex}
%{\sc JEL-Classification}: G12, G11, D53
%\end{verse}

\vfill
\end{titlepage}
\renewcommand{\thefootnote}{\arabic{footnote}}
\pagenumbering{arabic} \pagestyle{plain}

%\begin{center}
%\Large \textbf{\papertitle}
%\end{center}

%\vspace{0.5cm}

\section{Introduction}
We consider an incomplete continuous-time securities market with uncertainty generated by Brownian motions, which allows us to derive closed-form solutions for the equilibrium interest rate and market price of risk processes. The economy has a finite number of heterogeneous exponential utility investors, who receive partially unspanned income. The closed-form solutions facilitate a direct investigation of the impact of stochastic income volatility and preference heterogeneity on the resulting equilibrium interest rates and risk premia in a setting with unspanned income risk both at the individual and aggregate level.

The investors can trade continuously on a finite time-interval in a money market account and a single risky security, and they maximize expected time-additive exponential utility of continuous consumption. We  show that the impact of unspanned income risk on the risk premium compared to the Pareto efficient analogue depends on how risk premia are measured (instantaneously or discretely). We show, in a model-free manner, that unspanned income risk can never affect the equilibrium instantaneous market price of risk process in a setting with exponential utility investors and continuous income processes governed by Brownian motions. On the other hand, if risk premia are measured over finite time-intervals (as in empirical asset pricing studies), we show that unspanned income with countercyclical income volatility can increase the discrete market price of risk process compared to the Pareto efficient equilibrium in an otherwise identical complete market setting.

% , with discrete consumption and countercyclical income volatility, unspanned income can affect both the interest rate and the instantaneous the risk premium. Subsequently, we show, in a model-free manner, that with continuous consumption, unspanned income risk can never affect the equilibrium instantaneous risk premium. Lastly, even with continuous consumption, if risk premia are measured over finite time-intervals (as in empirical studies of asset pricing puzzles), we show that unspanned income with countercyclical income volatility can increase the equilibrium risk premium.

It is well-known that individual unspanned income risks lower the equilibrium interest rate compared to the Pareto efficient equilibrium in an otherwise identical complete market setting (see, e.g., \citeNP{Wan03}, \citeNP{KL10}, and \citeNP{CLM10}). This is due to the inefficient sharing of these risks and the resulting increased demand for precautionary savings. The key contribution of our paper is to demonstrate the non-trivial effects stochastic volatility of unspanned income can have on the resulting incomplete market equilibrium compared to the complete market analogue. While the equilibrium interest rate is always reduced, we show that the impact of unspanned income risk on the discrete market price of risk depends on whether the stochastic income volatility co-varies negatively or positively with the aggregate spanned income risk. The empirical evidence reported in, for example, \citeN{BL09} and \citeN{BFJ10} suggests that unspanned income risks are countercyclical, i.e., investors view their future income prospects as more uncertain in economic downturns. We show that unspanned income risk can increase (decrease) the discrete market price of risk process compared to the complete market setting if the stochastic income volatility is countercyclical (procyclical). Furthermore, we show in numerical examples that these impacts can be substantial.

\newpage

\shortciteN{CLM10} show that unspanned income risk has no impact on risk premia for aggregate spanned income risk (measured instantaneously or discretely) in settings with deterministic income volatility and exponential utility investors. As noted above, the introduction of stochastic income volatility does not change this result as long as returns and risk premia are measured instantaneously. This result follows from the fact that even when income volatility is stochastic, unspanned income risk is deterministic instantaneously. On the other hand, stochastic income volatility affects the return distributions over finite time-intervals and, thus, there can be an impact of unspanned income risk on the discrete market price of risk process. Increases in unspanned income risk affect an investor's expected utility negatively and, thus, if unspanned income volatility co-varies negatively with the aggregate spanned income risk, then the discrete market price of spanned income risk increases---negative shocks to spanned aggregate income do not only reduce contemporaneous consumption but these shocks also increase the volatility of subsequent unspanned income.

The questions of existence and characterization of complete market equilibria in continuous time and state models are well-studied.\footnote{See, e.g., Chapter 4 in \citeN{KS98} and Chapter 10 in \citeN{Duf01} for an overview of this literature. More recent references on complete market equilibria include \citeN{Zit06}, \citeN{CJMN09}, \citeN{AR08}, and \citeN{HMT09}.} The most common technique applied is based on the martingale method from \citeN{KLS87} and \citeN{CH89}, which in complete market settings provides an explicit characterization of the investor's optimizer. By using the so-called representative agent method, the search for a complete market equilibrium can be reduced to a finite-dimensional fixed-point problem. To the best of our knowledge, only \citeN{CH94}, \citeN{BC98}, \citeN{Zit10}, and \shortciteN{CLM10} consider the existence and characterization of a non-Pareto efficient equilibrium in a continuous-time trading setting.

Our setting is similar to that of \shortciteN{CLM10}, who derive closed-form solutions for all the equilibrium quantities in an economy with a finite number of heterogeneous exponential utility investors, and dividends and unspanned income governed by arithmetic Brownian motions. The crucial difference between the model in \shortciteN{CLM10} and our model is that we allow for stochastic income volatility and, still, we provide a tractable incomplete markets model for which the equilibrium price processes can be computed explicitly. Consequently, we can quantify the impact of the market incompleteness in the more realistic setting of stochastic income volatility supported by the empirical evidence (see \emph{op. cit.} \citeNP{BL09} and \shortciteNP{BFJ10}). The stochastic income volatility is a necessary ingredient in order to obtain an impact of unspanned income on the discrete market price of risk process. We incorporate a stochastic volatility \`a la Heston's model into the income and equilibrium risky security price dynamics. We derive explicit expressions for the equilibrium instantaneous and zero-coupon interest rates as well as for the instantaneous and discrete market price of risk processes in terms of the individual income dynamics and the absolute risk aversion coefficients. The resulting type of the equilibrium market price of risk process has been widely used in various optimal consumption-portfolio models (see, e.g., \citeNP{CV05} and \citeNP{Kra05}), whereas the resulting equilibrium interest rate  is similar to the celebrated CIR term structure model.

Translation invariant utility models (such as the exponential utility model we consider) allow consumption to be negative (see, e.g., the discussion in the textbook \citeNP{Ski09}). \citeN{SS05} show that this class of models is fairly tractable even when income is unspanned. We first conjecture the form of the equilibrium market price of risk process, and then we use the idea in \citeN{CH94} to re-write the individual investors' consumption-portfolio problems as problems with spanned income but heterogeneous beliefs. In certain affine settings with a deterministic interest rate, the exponential investor's value function is available in closed-form (see, e.g., \citeNP{Hen05}, \citeNP{Wan04}, \citeNP{Wan06}, and \shortciteNP{CLM10}). However, the incorporation of stochastic income volatility necessarily produces a stochastic equilibrium interest rate preventing the corresponding HJB-equation from having the usual exponential affine form. Therefore, the individual investor's value function is not available in closed-form in our setting. However, by using martingale methods, we obtain tractable expressions for the individually optimal consumption policies, which in turn are sufficient to produce the incomplete market equilibrium price processes.

In a discrete infinite time horizon model with a continuum of identical exponential utility investors, \citeN{Wan03} illustrates the negative impact unspanned  income risk can have on the equilibrium interest rate. Similarly, in a discrete-time setting, \citeN{KL10} provide sufficient conditions in a setting with a continuum of identical power utility investors under which unspanned idiosyncratic income risk will lower the equilibrium interest rate, but not affect the risk premium. \shortciteN{CLM10} present a continuous-time model with a finite number of exponential utility investors exhibiting the same interest rate phenomena, but also with no impact of unspanned income on the instantaneous risk premium. We extend these results by showing that as long as the income/consumption dynamics are continuous over time and the uncertainty is governed by Brownian motions, any equilibrium based on exponential preferences produces the same instantaneous risk premium as the standard Pareto efficient analogue. On the other hand, as noted above, we also prove that the discrete market price of risk process can be increased due to unspanned income risk if there is stochastic countercyclical income volatility.

\citeN{CD96}, and various extensions including \citeN{STY07}, produce similar equilibrium implications for the impact of unspanned income risk on interest rates and risk premia. They rely on a discrete-time analysis and a continuum of identical power utility investors with idiosyncratic income risks which wash-out at the aggregate level using a law of large numbers.  Given virtually any pattern of risky securities and bond prices, \citeN{CD96} show that individual income processes can be derived so that the (no-trade) equilibrium is consistent with these prices. In particular, if the cross-sectional volatility of the individual investors' income growth is countercyclical and sufficiently large, the model can produce equilibrium prices consistent with the high observed equity premium. In Chapter 21 in \citeN{Coc05} it is argued that cross-sectional income data do not show such large dispersion. In contrast to \citeN{CD96}, we consider a finite number of heterogeneous investors such that there is unspanned income risk both at the individual and at the aggregate level. Importantly, while the countercyclical income volatility in \citeN{CD96} pertains to the cross-sectional income distribution, the countercyclical income volatility in our model pertains to the individual investors' unspanned income risk.

Models based on a continuum of agents, such as \citeN{CD96} and \citeN{KL10}, rely on market clearing conditions defined by reference to a law of large numbers. \citeN{Jud85} and \citeN{Uhl96} discuss both technical and interpretation issues related to using such averaging market clearing conditions. Our model uses a finite number of investors and our market clearing conditions are required to hold pointwise, i.e., the realized aggregate demands are required to equal the aggregate supplies in equilibrium.

The paper is organized as follows. The next section introduces the structure of the economy in terms of the exogenously given quantities, and in terms of conjectures for the equilibrium price processes. Section 3 presents the investors' consumption-portfolio problems in which the investors take the conjectured price processes as given. Our main Section 4 first defines and then shows the existence of an equilibrium consistent with the conjectured price processes and, secondly, it examines the impact of market incompleteness on the equilibrium interest rates and risk premia. Section 5 re-states the investors' consumption-portfolio problems in the form of an equivalent complete market setting with heterogeneous beliefs. This allows us to derive the investors' equilibrium consumption processes explicitly, which is a key ingredient in the proof of the main equilibrium theorem stated in Section 4. The concluding Section 6 discusses variations of the model, and all proofs are in the appendix.

\section{Endowment and price processes}

We consider an endowment economy with a single non-storable consumption good which also serves as the num\'eraire, i.e., prices are quoted in terms of this good. The economy is populated by~$I<\infty$ consumer-investors all living on the time interval $[0,T]$, $T<\infty$. $(\Omega,\sF,\P)$ denotes the probability space on which all stochastic quantities are defined. $(W,Z)$ denotes an $1+I$ dimensional Brownian motion, where $W$ is scalar valued and $Z=(Z_i)_{i=1}^I$ is a vector of investor-specific Brownian motions. All Brownian motions $(W,Z_1,...,Z_I)$ are independent and the corresponding standard augmented Brownian filtration is denoted by $\sF_t$, $t\in[0,T]$. We consider $\sF:=\sF_T$ and we will often write $\E_t[\cdot]$ instead of $\E^\P[\cdot|\sF_t]$. %$\Lambda$ denotes the set of bounded real-valued functions on $[0,T]$, and
$\sL^p$ denotes the space of measurable and adapted processes $f$ such that
$$
\int_0^T |f_u|^pdu <\infty,\quad \P\text{-almost surely},\quad p\in\{1,2\}.
$$

\subsection{Exogenously specified quantities}

The investors have time-additive negative exponential utility of consumption with possible different degrees of absolute risk tolerance $\tau_i>0$, $i=1,...,I$. For simplicity, we assume that their time-preference rates are all equal to zero. Investor~$i$'s utility function over consumption is therefore
$$
U_i(x) := -e^{-x/\tau_i},\quad x\in\R, \quad i=1,...,I.
$$

The following process $v$ will be used to model stochastic income volatility. We define $v$ as the Feller process
\begin{align}\label{def:v}
dv_t &:=(\mu_v +\kappa_v v_t)dt + \sigma_{v}\sqrt{v_t}dW_t, \quad t\in[0,T],\quad v_0> 0,
\end{align}
where $\kappa_v,\mu_v,\sigma_v$ are constants such that $v$ remains strictly positive on $[0,T]$. Positivity is ensured by the first part of following assumption (Feller's condition).

\begin{assumption}\label{feller} The following two conditions are satisfied:
$$
\mu_v\ge \frac12\sigma_v^2>0.
$$
\end{assumption}

\noindent Investor $i$'s income is determined by the process
\begin{align}\label{def:Y_H}
dY_{it} &:= ( \mu_{Y_i} + \kappa_{Y_i} v_t )dt + \sqrt{v_t}\Big(\sigma_{Y_i}dW_t + \beta_{Y_i}dZ_{it}\Big),\quad Y_{i0}\in\R.
\end{align}
The parameters $(\mu_{Y_i},\kappa_{Y_i},\sigma_{Y_i},\beta_{Y_i})$, $i=1,...,I$, are  constants. The Brownian motion $W$ affects all investors' income processes, whereas the Brownian motion $Z_i$ models investor $i$'s idiosyncratic income risk. The income process $Y_i$ consists of the dividends from the investor's endowed portfolio of traded and non-traded assets with \emph{exogenous dividends} plus the investor's stream of labor income.

It is not immediate how to adjust our approach to cover the mean-reverting income models used in \citeN{Wan04} and \citeN{Wan06}. The affine optimal investment models used in \citeN{Wan04} and \citeN{Wan06} are based on an exogenously specified deterministic interest rate. However, the corresponding equilibrium interest rate cannot be deterministic or even independent of the investors' idiosyncratic income risk processes in these affine settings.\footnote{\citeN{Bre86} show that in complete markets settings the equilibrium interest rate is an increasing function of expected aggregate consumption growth. In mean-reverting income models the expected aggregate consumption growth depends on the level of aggregate consumption and, hence, the equilibrium interest rate is likely to depend on both the $W$-risk and on the investors' idiosyncratic risks $Z_i$. Our  income processes \eqref{def:Y_H} ensure that individual and aggregate income shocks are fully persistent. Therefore, the expected aggregate consumption growth is independent of the idiosyncratic income risk processes.} Unlike the power utility investor, stochastic interest rates complicate the exponential utility investor's optimal investment problem tremendously. As we shall see, the income processes \eqref{def:Y_H} produce a stochastic equilibrium interest rate, which is adapted to the filtration generated by $W$, and for which the individual exponential utility investor's  optimal investment problem remains partially tractable.

%The single risky security $S$ is taken to be an annuity, i.e., $S$ pays constant dividends at a unit rate. We take $S$ to be an annuity for simplicity, however, we could equally well consider a security paying out cumulative dividends at some rate $D_t$ satisfying certain properties, see the below discussion following Conjecture \ref{con:S_div}.

The aggregate income process $\sE_t := \sum_{i=1}^I Y_{it}$ has the dynamics
\begin{align}\label{def:sE}
d\sE_t = \Big(\mu_\sE +\kappa_\sE v_t\Big)dt + \sqrt{v_t}\Big( \sigma_\sE dW_t + \sum_{i=1}^I\beta_{Y_i} dZ_{it}\Big),\quad t\in [0,T],
\end{align}
where we have defined the constants
\begin{align}\label{sE_constants}
\tau_\Sigma := \sum_{i=1}^I \tau_i,\quad \sigma_\sE := \sum_{i=1}^I \sigma_{Y_i}, \quad \kappa_\sE :=  \sum_{i=1}^I\kappa_{Y_i}, \quad \mu_\sE :=  \sum_{i=1}^I\mu_{Y_i}.
\end{align}
In order to make the following discussions and interpretations unambiguous, we assume that $(\sigma_{Y_i},\beta_{Y_i})$ are nonnegative, $i=1,...,I$. The cross quadratic variation between the aggregate income process and the stochastic income volatility, i.e., $d\langle \sE_\cdot,v_\cdot\rangle_t= \sigma_v\sigma_\sE v_tdt$, is controlled by the parameter $\sigma_v$. In what follows $\sigma_v$ plays an important role, and we allow for both countercyclical ($\sigma_v<0$) and procyclical ($\sigma_v>0$) stochastic income volatility.

As we noted in the Introduction, \shortciteN{BFJ10} demonstrate empirically that income uncertainty is strongly countercyclical both at the aggregate, the firm, and the individual level and, hence, $\sigma_v<0$. Moreover, \citeN{BL09} and \shortciteN{BFJ10} demonstrate that income growth is negatively impacted by increases in the income volatility, for example, due to a ``higher value of waiting to invest" with non-convex capital adjustment costs and, hence, $\kappa_\sE<0$. In turn, this implies that the constant part of expected aggregate income growth must be positive, i.e., $\mu_\sE>0$, in order to have positive expected aggregate income growth (on average). In addition, \citeN{BL09} demonstrates that expected income growth rebounds following positive shocks to volatility. This is in our model captured by assuming that the volatility process is mean-reverting, i.e., $\kappa_v<0$. Therefore, in the following, the ``empirically relevant setting" refers to the parameter configuration:
$$
\mu_\sE>0,\quad \kappa_\sE<0,\quad \sigma_v<0,\quad \kappa_v<0.
$$

\subsection{Endogenously determined quantities}

The investors can trade continuously on the time interval $[0,T]$ in a money market account with price process $S^{(0)}$ and a single risky security with price process $S$. We begin with the money market account.

\begin{conjecture}\label{con:S0} The equilibrium price of the money market account has the dynamics
\begin{align}\label{eq:S0}
dS^{(0)}_t & = S^{(0)}_tr_tdt, \quad t\in[0,T],\quad S^{(0)}_0 =1,
\end{align}
where the $\sF_t^W:=\sigma(W_u)_{u\in[0,t]}$-adapted process $r$ is defined by
\begin{align}
r_t &:= \frac1{\tau_\Sigma}\Big\{\mu_\sE+ \Big(\kappa_\sE  -\frac12\sum_{i=1}^I \frac{\beta_{Y_i}^2}{\tau_i} -\frac{\sigma_{\sE}^2}{2\tau_\Sigma} \Big)v_t\Big\}.\label{r}
\end{align}
\end{conjecture}

\noindent For concreteness, we let the single risky security be an annuity paying out a unit dividend rate over $[0,T]$. We make the following conjecture.

\begin{conjecture}\label{con:S} There exists an $\sF_t^W:=\sigma(W_u)_{u\in[0,t]}$-adapted process $\sigma_S\in\sL^2$ with $\sigma_{St}\neq0$ for $t\in[0,T)$ such that the equilibrium price of the risky security has the dynamics
\begin{align}\label{eq:S}
dS_t + dt & = \Big(r_tS_t + \sigma_{St}\mu_S\sqrt{v_t}\Big)dt + \sigma_{St}dW_t,\quad S_0 >0,
\end{align}
where $r$ is defined by \eqref{r}, and the constant $\mu_S$ is defined by
\begin{align}
\mu_S := \frac{\sigma_{\sE}}{\tau_\Sigma}.\label{mu_S}
\end{align}
\end{conjecture}

\noindent The idiosyncratic Brownian motions $(Z_i)_{i=1}^I$ do neither appear directly in the risky security price dynamics \eqref{eq:S} nor in the spot interest rate dynamics \eqref{r}. Nevertheless, a key point of this paper is to explicitly quantify the significant impact the presence of the idiosyncratic unspanned risks $(Z_i)_{i=1}^I$ can have on $(S,S^{(0)})$. As will become clear, this impact is due to the coefficient $\frac12\sum_{i=1}^I {\beta_{Y_i}^2}/{\tau_i}$ in the interest rate dynamics \eqref{r}. %Since $d\langle \sE,v\rangle_t= v_t\sigma_v\sigma_\sE dt$, the parameter $\sigma_v$ controls the cross variation between the aggregate income process and the stochastic income volatility. In what follows $\sigma_v$ plays an important role, %, and to be consistent with empirical evidence this parameter should be negative (see, e.g., the discussions in Chapter 5 in \citeN{Gat06}), i.e., the stochastic income volatility is countercyclical. However,
%and we allow for both countercyclical ($\sigma_v<0$) and procyclical ($\sigma_v>0$) stochastic income volatility.

In order to state the third and final property regarding $(S,S^{(0)})$, we need the standard concept of state-price densities (see, e.g., Section 6F in \citeNP{Duf01}).  For clarity, we define these processes explicitly.

\begin{definition}\label{def:state_price} A \emph{local state-price density} $\xi^\nu$ has the form
$$
\xi^\nu_t = M_t^\nu/S^{(0)}_t,\quad t\in[0,T],\quad \xi^\nu_0=1,
$$
where $\nu\in \sL^2$, $W^\perp$ is a $W$-independent Brownian motion, and
$$
M_t^\nu := \exp\left(-\mu_S\int_0^t \sqrt{v_u} dW_u - \int_0^t \nu_u dW_u^\perp -\frac12\int_0^t \Big(\mu^2_Sv_u + \nu_u^2\Big)du\right).
$$
If, in addition, $\E[M_T^\nu]=1$, we call $\xi^\nu$ a \emph{state-price density}.
\end{definition}
\noindent The main property of local state-price densities is that both $\xi_t S_t^{(0)}$ and $\xi_t S_t$ are driftless under $\P$. For $\nu\in\sL^2$, $M^\nu$ is always a supermartingale with $\E[M_T^\nu] \le 1$. However, for $\xi^\nu$ to be a state-price density, we require $\nu\in\sL^2$ to produce the martingale property of $M^\nu$.

We will need the minimal state-price density $\xi^\text{min}$ for which $\nu:=0$, i.e.,
\begin{align}\label{def:min_state}
d\xi^\text{min}_t := -\xi^\text{min}_t\Big( r_t dt + \mu_S\sqrt{v_t} dW_t\Big),\quad \xi^\text{min}_{0}:=1.
\end{align}
The corresponding minimal martingale measure $\Q^\text{min}$ is defined via the Radon-Nikodym derivative on $\sF_T$ as (see, e.g., the survey \citeNP{FS10})
$$
dM^\text{min}_t := -M_t^\text{min}\mu_S\sqrt{v_t} dW_t, \quad M_0^\text{min}:=1, \quad \frac{d\Q^\text{min}}{d\P} := M_T^\text{min}>0.
$$
Since $v$ is a Feller process and $\mu_S$ is a constant, we see that Novikov's condition is satisfied, which in turn ensures that $M^\text{min}$ is a martingale.\footnote{\label{footnote:Novikov}More specifically, since $v_t$ is non-centrally $\chi^2$-distributed Novikov's condition is satisfied on small intervals. We can then use a localization argument like Example 3 on p.233 in \citeN{LS77} to obtain the global martingale property. We will use this observation multiple times in what follows.} Therefore, $\xi^\text{min}$ is indeed a state-price density and not just a local state-price density. Consequently, Girsanov's theorem ensures that
$$
dW_t^{\Q^\text{min}}:= dW_t + \mu_S\sqrt{v_t}dt,\quad W_0^{\Q^\text{min}}:=0,
$$
is a Brownian motion under $\Q^\text{min}$ which is independent of $(Z_1,...,Z_I)$.

\begin{definition}\label{def:impr} The \emph{instantaneous market price of risk process} for the Brownian motion $W$ is defined to be $\mu_S\sqrt{v_t}$ for $t\in[0,T]$ with $\mu_S$ defined by \eqref{mu_S}.
\end{definition}
%\footnote{It follows from \eqref{eq:S} that $\mu_S\sqrt{v_t}$ is equal to the instantaneous Sharpe ratio for $W$-risk.}

\noindent Since $d\langle \sE_\cdot,\mu_S\sqrt{v_\cdot}\rangle_t=\frac12\sigma_v\sigma_\sE\mu_S \sqrt{v_t}dt$, we see from Definition \ref{def:impr} that the instantaneous market price of risk process is countercyclical if, and only if, the income volatility is countercyclical, i.e., if, and only if, $\sigma_v<0$.

The following conjecture identifies the risky security by identifying the volatility process $\sigma_S$ appearing in price dynamics \eqref{eq:S}.

\begin{conjecture}\label{con:S_div} The equilibrium price of the risky security has the representation (note that $S_T=0$)
\begin{align}\label{def:S}
S_t = \E^{\Q^\text{min}}_t\left[\int_t^Te^{-\int_t^U r_sds} dU \right],\quad t\in[0,T].
\end{align}
\end{conjecture}

\noindent At a first glance, it may seem restrictive to take the single risky security to be an annuity. However, we can let the risky security be any security paying out dividends at rate $\delta_t$ as long as the process $\delta$ satisfies the following two properties:\footnote{One advantage of choosing the annuity as the risky security is that its stochastic return is only affected by changes in the stochastic volatility $v$, and not by changes in aggregate income $\sE$ (recall that the aggregate income shocks are fully persistent). It is this property of the annuity which allows us to demonstrate that the equilibrium $\sigma_{St}$ is non-zero on $[0,T)$.}
\begin{enumerate}
\item $\delta_t$ is an It\^o-process adapted to the filtration $\sF_t^W := \sigma(W_u)_{u\in[0,t]}$.
\item The following process is well-defined
$$
\E^{\Q^\text{min}}_t\left[\int_t^Te^{-\int_t^U r_sds} \delta_UdU\right],\quad t\in[0,T],
$$
and the $dW$-coefficient in these dynamics is non-zero on $[0,T)$.
\end{enumerate}
%The Heston-type income volatility \eqref{def:v} ultimately leads to CIR-type zero-cupon bond prices which can be explicitly computed. %This allows us to  verify that $\sigma_{St}>0$ for $t\in[0,T)$.

\noindent The second requirement is related to endogenous dynamic market completeness. \citeN{DH85}, \citeN{AR08} and \citeN{HMT09} provide conditions on the primitives of the economy under which an Arrow-Debreu equilibrium can be implemented by dynamic trading. In our setting, these conditions amount to ensuring that the $dW$-coefficient does not vanish in the above conditional expectation involving the $\sF_t^W$-adapted dividends $\delta_t$.

In the continuous-time securities market $(S^{(0)},S)$ with $\sigma_{St}\neq0$ for $t\in[0,T)$ (cf. Conjecture \ref{con:S}), all European claims written on the risky security, i.e., claims paying out $g(S_T)$ at time $T$ for some bounded payoff function $g$, are replicable.\footnote{The same also holds if $g$ is a bounded path functional of $(W_t)_{t\in[0,T]}$.} Hence, the assumption of only a single traded risky security with $\sF_t^W$-adapted dividends is not restrictive. The key incompleteness property is that the individual investor's income process $Y_{i}$ cannot be fully hedged due to the presence of $Z_i$ in the dynamics of $Y_i$. Therefore, $(S^{(0)},S)$ constitutes an incomplete continuous-time securities market. Consequently,  the standard method of describing the equilibrium by an representative agent cannot be applied.

%Finally, in the continuous-time securities market $(S^{(0)},S)$, all European claims written on the risky security, i.e., claims paying out $g(S_T)$ at time $T$ for some bounded payoff function $g$, are replicable.\footnote{The same also holds if $g$ is a bounded path functional of $(W_t)_{t\in[0,T]}$.} Hence, the assumption of only a single traded risky security with $\sF_t^W$-adapted dividends is not restrictive. The key incompleteness property is that the individual investor's income process $Y_{i}$ cannot be fully hedged due to the presence of $Z_i$ in the dynamics of $Y_i$. Therefore, $(S^{(0)},S)$ constitutes an incomplete continuous-time securities market. Consequently,  the standard method of describing the equilibrium by an representative agent  cannot be applied.

%The aggregate dividend payments made by the unit-supply risky security up to time $t$ equals $\int_0^t \delta_udu$ where $\delta_t$ denotes the dividend-rate process
%$$
%d\delta_t := \Big(\mu_\delta + \kappa_\delta v_t\Big)dt + \sigma_\delta  \sqrt{v_t}dW_t, \quad t\in[0,T],\quad \delta_0\ge0.
%$$
%Here $\sigma_\delta>0$ is a constant, $\mu_\delta,\kappa_\delta\in\R$, and $v$ is defined by \eqref{def:v}.

\section{The individual investor's problem}

Investor $i$ chooses trading strategies $(\theta^{(0)},\theta)$ as well as some consumption rate process $c$ in excess of the income $Y_i$. $\theta_t$ denotes the number of units held of the risky security in addition to the endowed portfolio of this asset (the endowed portfolio has dividends included in $Y_i$). Since the money market account has endogenous dividends paid at time $T$, the dividends generated by the endowed portfolio $\theta_{i0-}^{(0)}$ of this asset are \emph{not} included in the investor's income process $Y_i$. Therefore, $\theta_t^{(0)}$ denotes the total number of units held of the money market account at time $t\in[0,T]$. Consequently, $X^{\theta,c}_{it}:= \theta^{(0)}_t S^{(0)}_t + \theta_t S_t$ denotes the investor's total financial wealth (in addition to income) with initial condition $X_{i0}:=\theta_{i0-}^{(0)}S^{(0)}_0=\theta_{i0-}^{(0)}$. The self-financing condition becomes for $t\in[0,T]$
\begin{align}\label{def:wealth}
\begin{split}
X^{\theta,c}_{it} &= X_{i0} + \int_0^t \theta^{(0)}_u dS^{(0)}_u  + \int_0^t \theta_u (dS_u+du)  -\int_0^tc_udu\\
&=X_{i0} + \int_0^t r_uX_{iu}^{\theta,c} du + \int_0^t\theta_u\sigma_{Su}\Big(\mu_S\sqrt{v_u} du +dW_u\Big) -\int_0^tc_udu,
\end{split}
\end{align}
since the risky security is an annuity paying a unit dividend stream.
%{\bf Peter: I do not understand why the following is incorrect: Consider a discrete setting
%$$
%0 = t_0 < t_1 < .... < t_N = T.
%$$
%Since $S$ is measured after dividends are paid the budget restriction at time $t_n$ becomes
%$$
%\theta^{(0)}_{t_{n-1}} S^{(0)}_{t_n} + \theta_{t_{n-1}} (D_{t_n}-D_{t_{n-1}}+S_{t_n}) = \theta^{(0)}_{t_{n}} S^{(0)}_{t_n} + \theta_{t_n}S_{t_n}.
%$$
%We then have
%\begin{align*}
%&\theta^{(0)}_{t_{N}} S^{(0)}_{t_N} + \theta_{t_N}S_{t_N} - \Big(\theta^{(0)}_0 S^{(0)}_0 + \theta_0S_0\Big)\\&=
%\sum_{n=1}^N \theta^{(0)}_{t_{n}} S^{(0)}_{t_n} + \theta_{t_n}S_{t_n} - \Big(\theta^{(0)}_{t_{n-1}} S^{(0)}_{t_{n-1}} + \theta_{t_{n-1}}S_{t_{n-1}}\Big)\\
%&=\sum_{n=1}^N \theta^{(0)}_{t_{n-1}} S^{(0)}_{t_n} + \theta_{t_{n-1}} (D_{t_n}-D_{t_{n-1}}+S_{t_n}) - \Big(\theta^{(0)}_{t_{n-1}} S^{(0)}_{t_{n-1}} + \theta_{t_{n-1}}S_{t_{n-1}}\Big)\\
%&=\sum_{n=1}^N \theta^{(0)}_{t_{n-1}} (S^{(0)}_{t_n}-S^{(0)}_{t_{n-1}}) + \theta_{t_{n-1}} (D_{t_n}-D_{t_{n-1}}) - \theta_{t_{n-1}} (S_{t_{n}}-S_{t_{n-1}})\\
%&\to \int_0^T \theta^{(0)}_u dS^{(0)}_u + \int_0^T \theta_{u} \sigma_{Su}dW_u+ \int_0^T \theta_{u} \mu_S\sqrt{v_u}\sigma_{Su}du.
%\end{align*}
%}

In order to ensure well-posedness of the individual investor's optimization problem we need to impose conditions which ensure that the measurable and adapted processes $(\theta,c)$ are such that the wealth dynamics \eqref{def:wealth} are well-defined. Moreover, in order to rule out arbitrage, we need additional constraints on the possible choices. First, the investor is required to leave no obligations behind after the finite horizon:
\begin{align}\label{terminal_pos}
\P(X_{iT}^{\theta,c} \ge0 )=1.
\end{align}
Naturally, investor $i$ optimally chooses strategies $(\hat{\theta}_i,\hat{c}_i)$ such that $X_{iT}^{\hat{\theta}_i,\hat{c}_i} =0.$ We deem  $(\theta,c)$ admissible if additionally the process
\begin{align}\label{ineq:budget}
\xi^\nu_t X^{\theta,c}_{it} + \int_0^t\xi^\nu_u c_{u}du, \quad t\in[0,T],
\end{align}
is a supermartingale for all state-price densities $\xi^\nu$ (see Definition \ref{def:state_price}). In this case, we write $(c,\theta)\in\sA$. This supermartingale condition ensures that there are no arbitrage opportunities %\footnote{See p.123 in \citeN{Duf01} for the precise  definition of an arbitrage opportunity.}
in the admissible set $\sA$. In order to verify this claim, we let $\tau$ be a stopping time valued in $[0,T]$. Doob's optional sampling theorem produces
$$
\E\left[\xi^\nu_\tau X^{\theta,c}_{i\tau} + \int_0^\tau\xi^\nu_u c_{u}du\right] \le \xi_0^\nu X_{i0}=X_{i0}.
$$
By using this inequality with $c:=0$, we see that there are no arbitrage opportunities on $[0,T]$ in the admissible set $\sA$.

Investor $i$ maximizes time-additive expected utility stemming from consumption in addition to the investor's income, i.e., investor $i$ seeks $(\hat{c}_i,\hat{\theta}_i)\in\sA$ such that
\begin{equation}\label{eq:valuefunc}
%\begin{split}
\sup_{(c,\theta)\in \sA} \E\left[\int_0^T U_i(c_u+Y_{iu})du \right]\\
=\E\left[\int_0^T U_i(\hat{c}_{iu}+Y_{iu})du \right].
%\end{split}
\end{equation}

In Section 5 we show how to re-phrase \eqref{eq:valuefunc} in terms of heterogeneous beliefs and spanned income. This ultimately allows us to solve explicitly for the optimal consumption strategies $\hat{c}_i$ (see Theorem \ref{thm:ind} below),
$$
d\hat{c}_{it} =\Big\{\tau_i r_t + \Big(\frac12 {\tau_i\mu_S^2}+ \frac12 \frac{\beta_{Y_i}^2}{\tau_i}-\kappa_{Y_i} \Big)v_t - \mu_{Y_i}\Big\}dt+ \Big(\tau_i\mu_S - \sigma_{Y_i}\Big)\sqrt{v_t}dW_t,
$$
while only providing the abstract existence of $\hat{\theta}_i$ via the martingale representation theorem.

\section{Equilibrium}
%Both the money market account and the risky security are assumed to be in zero net-supply.
Before stating the following equilibrium definition (of the Radner-type), we recall that consumption $c_i$ is measured in excess of the income rates $Y_i$, and that the trading strategies $\theta_i$ denote the units held of the risky security \emph{in addition} to the investors' endowed portfolios of this asset. On the other hand, the trading strategies $\theta_i^{(0)}$ denote the total number of units held of the money market account. Since the money market account has endogenous dividends determined by the spot interest rates, this asset must be in zero net-supply in order to ensure that aggregate consumption is exogenous. Of course, this also implies that the endowments of the money market account must satisfy the clearing condition $\sum_{i=1}^I\theta_{i0-}^{(0)}=\sum_{i=1}^I X_{i0}=0$.

\begin{definition}\label{def:eq} An equilibrium is a set of security price processes $(S^{(0)},S)$, characterized by $(r,\mu_S)$, and a set of investor strategies $(\hat{c}_i,\hat{\theta}_i)\in\sA$ such that given $(r,\mu_S)$, the processes $(\hat{c}_i,\hat{\theta}_i)$ are optimal for investor $i$, $i=1,2...,I$, and such that all markets clear, i.e.,
\begin{align}\label{cond:clearing}
%\sum_{i=1}^IX^{\hat{c}_{i},\hat{\theta}_i}_{iT} = 1, \quad
\sum_{i=1}^I\hat{c}_{it}=0,\quad \sum_{i=1}^I\hat{\theta}_{it}=0,\quad\sum_{i=1}^I\hat{\theta}^{(0)}_{it}=0,\quad \P\otimes\text{Leb-a.e.}
\end{align}
%a minimal state-price density $\xi^{\text{min}}$ (see equation \eqref{def:min_state}) characterized by $(r,\mu_S)$ such that all markets clear, i.e.,
%\begin{align}\label{cond:clearing}
%\sum_{i=1}^IX^{\hat{c}_{i},\hat{\theta}_i}_{iT} = 1, \quad
%\sum_{i=1}^I\hat{c}_{it}=0,\quad \sum_{i=1}^I\hat{\theta}_{it}=0,\quad\sum_{i=1}^I\hat{\theta}^{(0)}_{it}=0,\quad \P\otimes\text{Leb-a.e.,}
%\end{align}
%and such that given $(r,\mu_S)$ the processes $(\hat{c}_i,\hat{\theta}_i)\in\sA$ are optimal for investor $i$, $i=1,2...,I$.
$\endproof$
\end{definition}

\begin{comment}\noindent The clearing conditions in \eqref{cond:clearing} reflect that we treat both $S^{(0)}$ and $S$ as financial securities. Hence, the unit dividend stream related to $S$ is not an exogenous consumption inflow into the economy. All conclusions in this paper remain valid if instead the annuity $S$ is considered a real security with an exogenous unit dividend stream, however, in that case the appropriate clearing conditions in \eqref{cond:clearing} become
$$
\sum_{i=1}^I\hat{c}_{it}=1,\quad \sum_{i=1}^I\hat{\theta}_{it}=1,\quad\sum_{i=1}^I\hat{\theta}^{(0)}_{it}=0,\quad \P\otimes\text{Leb-a.e.}
$$
\end{comment}

\noindent In order to state our main equilibrium existence theorem, we need the following assumption on the exogenous model parameters.

\begin{assumption}\label{ass:main} The parameters $\big(\kappa_v,\sigma_v, \kappa_\sE, \sigma_\sE, (\tau_i)_{i=1}^I, (\beta_{Y_i})_{i=1}^I\big)$ are such that the following two restrictions hold:
\begin{align}\label{ass:r_neg_restriction}
(\kappa_v -\frac{\sigma_\sE}{\tau_\Sigma}\sigma_v)^2 > 2\sigma_v^2\frac1{\tau_\Sigma}\Big( \sum_{i=1}^I \frac{\beta_{Y_i}^2}{2\tau_i} + \frac{\sigma_\sE^2}{2\tau_\Sigma} -\kappa_\sE\Big)\neq 0.
\end{align}
\end{assumption}

\noindent The first restriction in \eqref{ass:r_neg_restriction} trivially holds if
\begin{align}\label{ass:r_neg}
\frac{\kappa_\sE}{\tau_\Sigma} >  \frac1{\tau_\Sigma}\sum_{i=1}^I \frac{\beta_{Y_i}^2}{2\tau_i} + \frac{\sigma_\sE^2}{2\tau_\Sigma^2} .
\end{align}
When \eqref{ass:r_neg} holds, the spot interest rate process $r_t$ defined by \eqref{r} is bounded from below by $\mu_\sE/\tau_\Sigma$, i.e., $r_t$ is bounded from below by the constant part of risk-adjusted expected aggregate consumption growth per capita (recall we assume that the investors' time-preference rates are all equal to zero). Empirically, real interest rates can be negative and from \citeN{BL09} the constant $\kappa_\sE$ is likely to be negative. Therefore, we will use the weaker condition \eqref{ass:r_neg_restriction} in the following analysis. On the other hand, in the empirically relevant setting in which \eqref{ass:r_neg} fails, the spot interest rate process $r_t$ becomes unbounded from below and, consequently, zero-coupon bond prices may explode in finite time. Condition \eqref{ass:r_neg_restriction} ensures finite zero-coupon bond prices for all maturities which is all we need to prove our main equilibrium existence theorem.

The proof of the following main theorem shows that clearing in the good's market, i.e., $\sum_{i=1}^I\hat{c}_{it}=0$, ensures market clearing for both the risky security and the money market.

\begin{theorem} \label{thm:main} Under Assumptions \ref{feller} and \ref{ass:main}, %and $\sum_{i=1}^I X_{i0}=0$ Kasper: hvis vi skal have denne med så skal vi også have en tilsvarende med for det risiko fyldte aktiv
%we have that
%Assume that the explicit parameter restrictions
%\begin{align}
%\frac{\kappa_\delta +\kappa_\sE}{\tau_\Sigma}  &> \frac1{2\tau_\Sigma}\Big(\sum_{i=1}^Ia_i\beta_{Y_i}^2 +\frac{(\sigma_\delta+ \sigma_{\sE})^2}{\tau_\Sigma}\Big),\label{par_restriction}\\
%\kappa_v &< \frac{\sigma_\delta+ \sigma_{\sE}}{\tau_\Sigma}\sigma_v, \quad\
%\kappa_v^2 +2\sigma_v^2(\kappa_\sE+\kappa_\delta) &> \sigma_v^2\frac1{\tau_\Sigma} \Big( \sum_{i=1}^Ia_i\beta_{Y_i}^2 +2\kappa_v(\sigma_\sE+\sigma_\delta)  \Big)\label{par_restriction1}.
%\end{align}
the security price processes $(S^{(0)},S)$ characterized by $r$ and $\mu_S$ defined in \eqref{r} and \eqref{mu_S}, respectively, with the resulting individually optimal strategies $(\hat{c}_i,\hat{\theta}_i)\in\sA$, $i=1,2...,I$, constitute an equilibrium for which Conjectures \ref{con:S0}, \ref{con:S}, and \ref{con:S_div} hold.
\end{theorem}

\noindent The following lemma shows that our equilibrium produces exponential-affine zero-coupon bond prices, and this property constitutes an important ingredient in the proof of Theorem \ref{thm:main}. We refer to the appendix in \citeN{KO96} for a detailed description of Riccati equations, see, in particular, the discussion on \emph{normal} Riccati solutions.

\begin{lemma}\label{lem:ZCB} Under Assumption \ref{ass:main}, the following coupled system of ODEs with $a(0)=b(0)=0$ and for $s>0$
\begin{align}
b'(s) &= b(s)(\kappa_v-\frac{\sigma_\sE}{\tau_\Sigma}\sigma_v) +\frac12b(s)^2\sigma_v^2 +\frac1{\tau_\Sigma} \Big( \frac12\sum_{i=1}^I \frac{\beta_{Y_i}^2}{\tau_i} + \frac{\sigma_\sE^2}{2\tau_\Sigma} -\kappa_\sE  \Big),\label{ode:b}\\
a'(s) &= \frac{\mu_\sE}{\tau_\Sigma}-b(s)\mu_v,\label{ode:a}
\end{align}
has unique non-exploding solutions satisfying $b(s)\neq 0$ for $s\in(0,\infty)$. Furthermore, for $\mu_S$ defined by \eqref{mu_S}, we have for $ t\in[0,U]$ that the zero-coupon bond prices are given by
\begin{align}\label{def:zcb}
B(t,U) := \E^{\Q^\text{min}}_t[ e^{-\int_t^U r_s ds}]=\exp\Big(b(U-t)v_t-a(U-t)\Big).
\end{align}
\end{lemma}

\noindent Depending on whether \eqref{ass:r_neg} holds, the second restriction in \eqref{ass:r_neg_restriction} ensures that \eqref{ode:b} has a positive or negative solution $b(s)$ for $s\in[0,\infty)$. In the empirically relevant setting in which \eqref{ass:r_neg} fails, the solution to \eqref{ode:b} is positive. Therefore, the zero-coupon bond prices are increasing in the volatility $v_t$, which is consistent with increasing incentives for precautionary savings when income risk increases (see, e.g., the discussion in \shortciteNP{CLM10}).

The proof of Theorem \ref{thm:main} shows that Conjecture \ref{con:S_div} holds with the volatility coefficient
\begin{align}\label{eq:vol_coefficient}
\sigma_{St} := \sigma_v\sqrt{v_t} \int_t^T B(t,U)b(U-t)dU, \quad t\in[0,T),
\end{align}
which is non-zero on the interval $[0,T)$ under Assumption \ref{ass:main}. The sign of $\sigma_S$ is determined by the sign of $\sigma_v$ and the sign of the function $b$. In the empirically relevant setting in which the income volatility is countercyclical ($\sigma_v<0$), and in which increasing income volatility reduces the expected aggregate income growth ($\kappa_\sE<0$, implying that \eqref{ass:r_neg} fails), the function $b$ is positive. This implies that the volatility process \eqref{eq:vol_coefficient} is negative. Therefore, the instantaneous risk premium for the annuity, i.e., $\sigma_{St}\mu_S\sqrt{v_t}/S_t$, is also negative. %Of course, this result reflects that the annuity hedges the spanned aggregate $W$-risk in this setting.

Finally, we mention that Theorem \ref{thm:main} does not make any  uniqueness statement regarding the equilibrium. In other words, we are \emph{not} claiming that $S^{(0)}$ defined by \eqref{eq:S0} and $S$ defined by \eqref{eq:S} is the only equilibrium possible in our pure exchange economy.

\subsection{Equilibrium impacts due to incompleteness}

In this section we analytically show how the incomplete market equilibrium established in Theorem \ref{thm:main} can be used to simultaneously explain the risk-free interest rate puzzle and the equity premium puzzle. We compare the equilibrium characterized in Theorem \ref{thm:main} to the equilibrium in an otherwise identical complete market economy in which all risks are spanned.

In the complete market economy, there exists a representative agent, and the equilibrium is characterized by the representative agent's first-order condition. The representative agent is modeled by the utility function
$$
U_\text{rep}(x;\gamma) := \sup_{\sum_{i=1}^I x_i = x} \sum_{i=1}^I \gamma_iU_i(x_i),\quad \gamma \in \R^I_+,\quad x\in \R,
$$
where $\gamma$ is a  Negishi-weight vector. Since each investor is modeled by a negative exponential utility function, the representative agent's utility function becomes (see, e.g., Section 5.26 in \citeNP{HL88})
$$
U_\text{rep}(x;\gamma) = -e^{-\frac1{\tau_\Sigma} x}\;\prod_{i=1}^I (\frac{\gamma_i}{\tau_i})^{\frac{\tau_i}{\tau_\Sigma}},\quad x\in \R.
$$
This expression shows that the weight $\gamma$ does not matter for the representative agent's preferences (Gorman aggregation). The first-order condition for the representative agent produces the proportionality requirement
\begin{align}\label{rep_prop}
e^{-\frac1{\tau_\Sigma} \sE_t} \propto  \xi_t^\text{rep}, \quad t\in [0,T],
\end{align}
where the aggregate income process $\sE_t$ is defined by \eqref{def:sE}, and $\xi_t^\text{rep}$ is the unique state-price density in the representative agent setting. By computing the dynamics of both sides of \eqref{rep_prop} and matching the coefficients we find the spot interest rate based on the representative agent economy to be
\begin{align}\label{repr_agent_r}
r^\text{rep}_t := \frac1{\tau_\Sigma}\mu_\sE  + \frac1{\tau_\Sigma} \Big(\kappa_\sE-\frac1{2\tau_\Sigma}\Big\{\sum_{i=1}^I \beta_{Y_i}^2 + \sigma_\sE^2 \Big\}\Big)v_t, \quad t\in[0,T].
\end{align}
Since $\tau_\Sigma := \sum_{i=1}^I \tau_i$, we have that $\tau_\Sigma \ge \tau_i$ for all $i$, which produces the key inequality
\begin{align}\label{ineq:beta}
\sum_{i=1}^I \frac{\beta_{Y_i}^2}{\tau_i} \ge \frac1{\tau_\Sigma}\sum_{i=1}^I \beta_{Y_i}^2.
\end{align}
In an economy with unspanned idiosyncratic risks $Z_i$, this inequality combined with Theorem \ref{thm:main} produces the interest rate reduction
\begin{align}\label{rate_diff}
r^\text{rep}_t - r_t = \frac1{2\tau_\Sigma}\left( \sum_{i=1}^I \frac{\beta_{Y_i}^2}{\tau_i} -\frac1{\tau_\Sigma}\sum_{i=1}^I \beta_{Y_i}^2\right)v_t\ge0,
\end{align}
which is an analogue of the result presented in \shortciteN{CLM10} (compare to their equation (30)) although the interest rate reduction in our model is stochastic due to the common stochastic income volatility $v$.

Similarly, from the dynamics of \eqref{rep_prop} we find that the instantaneous market price of risk process based on the  representative agent is identical to the market price of risk process derived in Theorem \ref{thm:main}, namely $\mu_S\sqrt{v_t} = \frac{\sigma_\sE}{\tau_\Sigma}\sqrt{v_t}$. This is also an analogue of the result presented in \shortciteN{CLM10} (compare to their equation (27)). This equilibrium implication is not limited to our particular income model \eqref{def:v}-\eqref{def:Y_H}. Theorem \ref{thm:cont} below shows that this result holds true in \emph{any} model based on exponential investors and continuous income rates based on It\^o-processes driven by Brownian motions.

%\subsection{Risk premia measured over finite time-intervals}
We will next establish that unspanned income risk with stochastic volatility can affect the risk premium measured over finite time-intervals, even though there is no impact on the instantaneous market price of risk process as demonstrated above. Our motivation is that empirical studies of asset pricing properties, such as the risk-free rate and the equity premium puzzles, necessarily must measure returns, spot interest rates, and risk premia over finite time-intervals, where the length $U>0$ of the time-intervals is determined by the sampling frequency. In order to precisely quantify risk premia measured over $[0,U]$, we need to introduce the minimal forward measure $\Q^U$. Since the equilibrium spot interest rate derived in Theorem \ref{thm:main} is stochastic, the minimal martingale measure $\Q^\text{min}$ and the minimal forward measure $\Q^U$ differ. %, and it is this difference which produces a non-trivial impact on equilibrium risk premia measured over $[0,U]$.
 The probability measure $\Q^U$ is defined by the Radon-Nikodym derivative on $\sF_U$ as
$$
\frac{d\Q^U}{d\Q^\text{min}}:= \frac{\exp\left(-\int_0^U r_udu\right)}{B(0,U)},\quad U\in(0,T].
$$
Lemma \ref{lem:ZCB} provides an explicit representation for equilibrium zero-coupon bond prices $B(t,U)$. Based on this lemma, Girsanov's theorem ensures that
$$
dW^{\Q^U}_t := dW^\text{min}_t - b(U-t)\sigma_v\sqrt{v_t}dt = dW_t + \Big(\mu_S- b(U-t)\sigma_v\Big)\sqrt{v_t}dt,
$$
is a $\Q^U$-Brownian motion, where the deterministic function $b$ is defined by the Riccati equation \eqref{ode:b}. We can then make the following definition.
\begin{definition}\label{def:premium} Under Assumption \ref{ass:main}: The \emph{discrete market price of risk process} measured over the finite time-interval $[0,U]$, $U\in(0,T]$, is defined by $\mu_S^\text{dis}(t)\sqrt{v_t}$, where%\footnote{For $\mu_S^\text{dis}(t)\sqrt{v_t}$ to be interpreted as a risk premium, we implicitly consider a risky security with a unit volatility parameter. In this section we are interested in examining the impact of unspanned income on the equity premium and, thus, we must consider a security which has positive cross quadratic variation with the spanned $W$-risk.}
$$
\mu_S^\text{dis}(t) := \mu_S- b(U-t)\sigma_v =  \frac{\sigma_\sE}{\tau_\Sigma} - b(U-t)\sigma_v,\quad t\in[0,U].
$$
\end{definition}

\noindent Our reasoning behind Definition \ref{def:premium} is the following. Let $\sigma_X\in\sL^2$ be a bounded process and consider a traded security with price process (use the wealth dynamics \eqref{def:wealth} with $c:=0$)
$$
dX_t := r_t X_t dt + \sigma_{Xt} dW^{\Q^\text{min}}_t,\quad t\in[0,T],\quad  X_0\in\R.
$$
The main characterizing property of $\Q^U$ is that all prices of traded securities denominated in terms of the price of the zero-coupon bond maturing at time $U$ have no drift under $\Q^U$. Since $\sigma_X$ is bounded, we therefore have
$$
\frac{X_0}{B(0,U)} = \E^{\Q^U}\Big[\frac{X_U}{B(U,U)}\Big] = \E^{\Q^U}[X_U],
$$
where the last equality follows from $B(U,U)=1$. This identity implies that the expected return over the interval $[0,U]$ under the minimal forward measure $\Q^U$ is equal to the zero-coupon rate for this interval, i.e.,
\begin{align}\label{Q_U_relation}
\E^{\Q^U}\Big[\frac{X_U-X_0}{X_0}\Big]=\frac{1-B(0,U)}{B(0,U)}.
%\\&= \frac1{X_0} \E^{\Q^U}\left[\int_0^U \Big(r_tX_t + \sigma_{Xt}\sigma_vb(U-t)\sqrt{v_t}\Big)dt\right].
\end{align}
In other words, the process $\mu_S^\text{dis}(t)\sqrt{v_t}$ is the drift-correction in the $W$-dynamics needed to produce the riskless return as the expected return under $\Q^U$ of $X$ over the interval $[0,U]$.

We focus on the discrete market price of risk process, since the $W$-drift correction $\mu_S^\text{dis}(t)\sqrt{v_t}$ is universal across all traded securities. Alternatively, we could consider the risk premium over the interval $[0,U]$ for a security with the price process $X_t$. This premium is defined by the difference
$$
\E\left[\frac{X_U-X_0}{X_0}\right]-\frac{1-B(0,U)}{B(0,U)} = - \frac1{X_0}\text{Cov}\left(\frac{d\Q^U}{d\P}, X_U\right),\quad U\in(0,T],
$$
where the equality follows from \eqref{Q_U_relation}. The Radon-Nikodym derivative $\frac{d\Q^U}{d\P}$ is completely determined by the discrete market price of risk process $\mu_S^\text{dis}(t)\sqrt{v_t}$ via
$$
\frac{d\Q^U}{d\P}:= M_U^{\Q^U},\quad dM_t^{\Q^U}:=-M_t^{\Q^U}\mu_S^\text{dis}(t)\sqrt{v_t}dW_t,\quad t\in[0,U], \quad M_0^{\Q^U}:=1.
$$
%From this relation we see that if there is an impact of unspanned income on the discrete market price of risk $\mu_S^\text{dis}(t)\sqrt{v_t}$, there will also be an impact on the risk premium over the interval $[0,U]$.
From this we see that the impact on the risk premium over the interval $[0,U]$ due to market incompleteness depends on the security. In other words, unlike the discrete market price of risk process, the significance of the impact on the risk premium over $[0,U]$ depends on the security's  volatility process $\sigma_{X}$. Moreover, contrary to when returns and risk premia are measured instantaneously, normalizing the risk premium over the interval $[0,U]$ by the standard deviation of the security's return $\frac1{X_0}\sqrt{\text{Var}[X_U]}$ to produce the discrete ``Sharpe ratio'' does not remove the dependence on the security's volatility process $\sigma_X$.

Similarly to the probability measures $\Q^\text{min}$ and $\Q^U$, we can introduce $\Q^\text{min}_\text{rep}$ and $\Q^U_\text{rep}$ corresponding to the representative agent based on the spot interest rate $r_t^\text{rep}$ defined by \eqref{repr_agent_r}. This interest rate $r_t^\text{rep}$ produces the zero-cupon bond prices for $0\le t\le U\le T$:
\begin{align}\label{zcb_rep}
B_\text{rep}(t,U):=\E^{\Q^\text{min}}_t[ e^{-\int_t^U r_s^\text{rep} ds}] = \exp\Big(b_\text{rep}(U-t)v_t -a_\text{rep}(U-t)\Big),
\end{align}
where $a_\text{rep}$ and $b_\text{rep}$ are defined by $a_\text{rep}(0)=b_\text{rep}(0) =0$ and for $t\in[0,T]$
\begin{align*}
b_\text{rep}'(t) &= b_\text{rep}(t)(\kappa_v-\mu_S\sigma_v) +\frac12b_\text{rep}(t)^2\sigma_v^2 +\frac1{\tau_\Sigma} \Big( \frac1{2\tau_\Sigma}\sum_{i=1}^I\beta_{Y_i}^2 +\frac{\sigma_\sE^2}{2\tau_\Sigma} -\kappa_\sE  \Big),\\
a_\text{rep}'(s) &= \frac{\mu_\sE}{\tau_\Sigma}-b_\text{rep}(s)\mu_v.
\end{align*}
By using the inequality \eqref{ineq:beta}, we see that Assumption \ref{ass:main} ensures that the Riccati equation describing $b_\text{rep}$ has a unique non-exploding solution on $[0,\infty)$. Therefore, for $U\in(0,T]$, the process
$$
dW^{\Q_\text{rep}^U}_t := dW^{\Q^\text{min}}_t - b_\text{rep}(U-t)\sigma_v\sqrt{v_t}dt= dW_t + \Big(\mu_S - b_\text{rep}(U-t)\sigma_v\Big)\sqrt{v_t}dt,
$$
is a Brownian motion under the representative agent's minimal forward measure $\Q_\text{rep}^U$.  The discrete market price of risk process measured over $[0,U]$ corresponding to the representative agent is defined similarly to Definition \ref{def:premium} as $\mu_S^\text{dis,rep}(t)\sqrt{v_t}$ where
$$
\mu_S^\text{dis,rep}(t) := \mu_S- b_\text{rep}(U-t)\sigma_v = \frac{\sigma_\sE}{\tau_\Sigma} - b_\text{rep}(U-t)\sigma_v,\quad t\in[0,U].
$$

By comparing the coefficients for the two Riccati equations describing $b$ and $b_\text{rep}$ and using the inequality \eqref{ineq:beta}, we see that $b_\text{rep}(t)\le b(t)$ for all $t\in[0,T)$. Consequently, provided that $\sigma_v\neq 0$ as in the second part of Assumption \ref{feller}, we obtain an impact on the equilibrium discrete market price of risk process measured over $[0,U]$. In particular, if the stochastic income volatility is countercyclical ($\sigma_v<0$), the equilibrium discrete market price of risk process measured over finite time-intervals is higher than in an otherwise identical complete market.

Similarly to the derivation of $\sigma_S$ in \eqref{eq:vol_coefficient} presented in the proof of Theorem \ref{thm:main}, we can show that the annuity's volatility coefficient in the representative agent setting is
$$
\sigma_{St}^\text{rep} := \sigma_v\sqrt{v_t}\int_t^T B_\text{rep}(t,U)b_\text{rep}(U-t)dU, \quad t\in[0,T].
$$
\noindent In the empirically relevant setting in which $\sigma_v<0$ and $\kappa_\sE<0$, we have $0\le b_\text{rep}(t)\le b(t)$ and, hence, also $a(t) \le a_\text{rep}(t)$, for all $t\in[0,T)$. It therefore follows from \eqref{def:zcb} and \eqref{zcb_rep}  that $B_\text{rep}(t,U)\leq B(t,U)$ which produces the inequality
$$
\sigma_{St} \le  \sigma_{St}^\text{rep} < 0,\quad t\in[0,T).
$$
In other words, the drift-correction $\sigma_{St}^\text{rep}\mu_S^\text{dis,rep}(t)\sqrt{v_t}$ in the $S$-dynamics needed to produce the riskless return as the expected return of the annuity over the interval $[0,U]$  in the complete market setting is larger (less negative) than in the incomplete market setting.

This section has explicitly illustrated the equilibrium impacts due to market incompleteness, which we summarize as follows:
\begin{enumerate}
\item The equilibrium spot interest rate is impacted negatively.
\item The equilibrium instantaneous market price of risk process is unaffected. Theorem \ref{thm:cont} below shows that this feature carries over to any model based on exponential utility investors and continuous income processes driven by Brownian motions.
\item The equilibrium discrete market price of risk process measured over a finite interval $[0,U]$ is impacted, and the sign of the impact depends on the sign of $\sigma_v$.
\item The equilibrium volatility coefficient of the risky security is impacted, and the sign of the impact depends on the sign of $\sigma_v$.
\end{enumerate}

\subsection{Numerical illustrations}
This section serves to illustrate that the impact on the equilibrium interest rate and the discrete market price of risk stemming from investors receiving partially unspanned income with stochastic volatility can be significant. The numerical values reported in this section only serve to illustrate the potential impact. %In order to calibrate our model to market data several modifications are needed and we refer to the discussion in the beginning of Section 5.

The impact on the interest rate and on the discrete market price of risk is determined by \eqref{ineq:beta}:
\begin{align*}
\Delta_\beta &:= \frac1{\tau_\Sigma}\sum_{i=1}^I \frac{\beta_{Y_i}^2}{\tau_i}-\frac1{\tau_\Sigma^2} \sum_{i=1}^I \beta_{Y_i}^2=\frac1{\tau_\Sigma^2} \sum_{i=1}^I \big (\frac{\tau_\Sigma}{\tau_i} - 1  \big ) \beta_{Y_i}^2\ge0.
\end{align*}

We consider first a homogeneous investor setting in which all investors have the same risk tolerance $\tau_i:=\tau$ as well as the same unspanned income risk parameter $\beta_{Y_i}:=\beta_{Y}$, $i=1,...,I$. In this setting, $\tau_\Sigma=I\tau$, and we find that
$$
\Delta_\beta = \frac1{\tau^2} \big (1- \frac1{I} \big ) \beta_{Y}^2 \uparrow \frac1{\tau^2} \beta_Y^2,
$$
as $I\to \infty$.
For given parameter values, Table \ref{table:1} shows the impact of unspanned income risk with countercyclical stochastic volatility on the interest rate [column 2] and on the discrete market price of risk measured over $[0,U]$ with initial condition $v_0:=1$  [column 3].
\begin{table}[h]
\begin{tabular}{cccc}
\hline\hline
$I$ & $r^\text{rep}_0-r_0$ & $\mu^\text{dis}_{S}(0)-\mu^\text{dis,rep}_{S}(0)$\\%\sigma_v\big(b(0)-b_{rep}(0)\big)$ \\
\hline
2   &0.0400&  0.0094  \\ % &0.0198\\
5   & 0.0640&  0.0151  \\ %& 0.0316 \\
10  & 0.0720& 0.0169   \\%&0.0356\\
100  &  0.0792&0.0186   \\%& 0.0391 \\
1000  &  0.0799&0.0188  \\%& 0.0395\\
$\infty$   &  0.0800&0.0188 \\ %& 0.0395 \\
\hline
\end{tabular}
\centering \caption{Equilibrium effects of increasing the number of investors $I$ for the volatility parameters $v_0:=1$, $\mu_v :=0.05$, $\kappa_v:=-0.7$, and $\sigma_v:=-0.3$. The investor parameters are $\tau_i := \tfrac{1}{2}$, $\beta_{Y_i} :=0.2$, $\kappa_{Y_i} := 0$, and $\sigma_{Y_i} := 0.3$ for all $i$. The horizon is $U:=1$.}
\label{table:1}
\end{table}

Secondly, we consider a heterogeneous investors setting in which we can split the population into two homogenous groups $A$ and $B$ with characteristics $(\tau_A,\beta_{Y_A})$ and $(\tau_B,\beta_{Y_B})$. The weight $w$ denotes group $A$'s proportion of the overall population. Table \ref{table:2} reports the increase in the discrete market price of risk measured over $[0,U]$, i.e., $\mu^\text{dis}_{S}(0)-\mu^\text{dis,rep}_{S}(0)$, for various combinations of risk tolerance parameters and population distributions in the limiting model $(I\to\infty)$. We see that the impact on the discrete market price of risk is highest when the less risk tolerant investors face the largest unspanned income risk.

\begin{table}[h]\setlength{\extrarowheight}{6pt}
\begin{tabular}{llllll}
\hline\hline
    & \multicolumn{4}{c}{$(\tau_A,\tau_B)$} \\\cline{2-5}
$w $ &$(\tfrac{1}{2},\tfrac{1}{2})$& $(\tfrac{1}{2},\tfrac{1}{3})$ & $(\tfrac{1}{3},\tfrac{1}{2})$ & $(\tfrac{1}{3},\tfrac{1}{3})$ \\
\hline
1.00 &    0.0047   &0.0047&  0.0111    &0.0111 \\
0.75 &   0.0223  & 0.0349 & 0.0332  & 0.0528\\
0.50 &    0.0400  &0.0720  & 0.0504  & 0.0946 \\
0.25 &  0.0577  &0.1186  &0.0642 &  0.1367 \\
0.00 & 0.0755 &0.1789 &0.0755 &  0.1789 \\
\hline
\end{tabular}
\centering \caption{Increase in the discrete market price of risk $\mu^\text{dis}_{S}(0)-\mu^\text{dis,rep}_{S}(0)$ in the limiting case $(I\to \infty)$ for various weights $w$ and various risk tolerance parameters $(\tau_A,\tau_B)$. The numbers are based on $\beta_{Y_A}:= 0.1$, $\beta_{Y_B}:=0.4$, whereas the remaining exogenous parameters are as in Table \ref{table:1}.}
\label{table:2}
\end{table}

\subsection{No impact on the instantaneous market price of risk process}
In this section we show that the instantaneous market price of risk process based on the representative agent is \emph{always} identical to the equilibrium instantaneous market price of risk process in a setting based on exponential investors and continuous income processes governed by Brownian motions. We consider the following model for $t\in[0,T]$:
\begin{equation}
\begin{split}\label{S:gen}
dS^{(0)}_t  &:= r_tS^{(0)}_tdt,\quad S_0^{(0)}:=1,\\
dS_t + \delta_tdt &:= \Big(r_tS_t + \lambda'_t\sigma_{St}'\Big)dt +\sigma_{St}'dB_t,\quad S_0 \in\R,
\end{split}
\end{equation}
for some $(\delta, r)\in\sL^1$, $(\sigma'_S,\lambda')\in\sL^2$, $\sigma'_S\neq 0$ and a Brownian motion $B$. %Both assets $(S^{(0)},S)$ are in zero net-supply.
In the following theorem we refer to Definition \ref{def:state_price} for the notion of a local  state-price density $\xi^\nu$.

\begin{theorem}\label{thm:cont} For $t\in[0,T]$ we consider the income  dynamics
\begin{align*}
%d\delta'_t &:= \mu'_{\delta t}dt + \sigma'_{\delta t} dB_t + \beta'_{\delta t} dB_{\delta t}^\perp,\quad \delta'_0\in\R\\
dY'_{it} &= \mu'_{Y_it}dt + \sigma'_{Y_it}dB_t + \beta'_{Y_it} dB^\perp_{it},\quad Y'_{i0}\in\R.
\end{align*}
Here $
%B_\delta^\perp,
B_1^\perp,...,B_I^\perp$ denote possible dependent one-dimensional Brownian motions independent of $B$, $
%\mu'_\delta,
\mu'_{Y_i}\in\sL^1$, and $
%(\sigma_\delta, \beta_\delta,
(\sigma'_{Y_i}, \beta'_{Y_i})\in\sL^2$.  Assume that \eqref{S:gen} constitutes an equilibrium in which each investor's optimal consumption process $\hat{c}_{it}$ satisfies the following first-order condition
\begin{align}\label{ind:FOC}
U_i'(\hat{c}_{it} + Y'_{it}) \propto \hat{\xi}_{it},\quad t\in[0,T],\quad i=1,...,I,
\end{align}
where $\hat{\xi}_{i}$ is an investor-specific local state-price density.  Then the equilibrium instantaneous market price of risk process $\lambda'$ satisfies $\lambda'_t = \frac1{\tau_\Sigma}\sum_{i=1}^I  \sigma'_{Y_it}$.
\end{theorem}

\noindent In the setting of this theorem, let $\sE'_t := \sum_{i=1}^I  Y'_{it}$ denote the aggregate endowment. By computing the dynamics of the representative agent's state-price density (proportional to $e^{-\frac1{\tau_\Sigma}\sE'_t}$), we see that the instantaneous market price of risk process based on the representative agent agrees with $\lambda'$ stated in Theorem \ref{thm:cont}. In other words, Theorem \ref{thm:cont} shows that any model based on exponential utility investors and continuous income processes governed by Brownian motions produces the same instantaneous market price of risk process as suggested by the standard representative agent model.

There is no loss of generality in assuming the above form for $(Y'_i)_{i=1}^I$ and $S$. Indeed, by assuming that an equilibrium risky security price $S$ exists, we can use L\'evy's characterization for Brownian motion as well as the martingale representation theorem for $\sF_t := \sigma(W_u,Z_{u1},...,Z_{uI})_{u\in[0,t]}$ to write the martingale component of $dS$ as $\sigma'_{St}dB_t$ for some Brownian motion $B$ and some process $\sigma'_S\in\sL^2$. Subsequently, we can decompose the martingale part of $Y_i'$ into its projection onto $B$ and some residual orthogonal martingale component (possibly depending on $i$) which produces the above form for $dY'_i$ for $i=1,...,I$.

Finally, we discuss the first-order condition \eqref{ind:FOC}. In the case of utility functions defined on the positive semi-axis, \citeN{CSW01} show that the introduction of unspanned endowments may require finite additive measures in the dual space, in which case \eqref{ind:FOC} makes no sense.  However, \citeN{OZ09} show that for utility functions defined over $\R$---such as our setting---the dual optimizer is always a (countable additive) measure and \eqref{ind:FOC} holds. Both papers \shortciteN{CSW01} and \citeN{OZ09} consider the case of expected utility of terminal wealth only and instead of re-proving \citeN{OZ09} to fit our case of continuous consumption, we have opted for assuming \eqref{ind:FOC} upfront.

\section{Heterogeneous beliefs}

This section contains the key ingredient required to prove our main Theorem \ref{thm:main}. We introduce a technique which allows us to partially solve the individual investor's consumption-portfolio problem \eqref{eq:valuefunc}. Because the interest rate $r_t$ is stochastic, the PDE produced by the HJB-approach does not have the usual exponential affine solution that \citeN{Hen05} and \shortciteN{CLM10} rely on. Inspired by \citeN{CH94}, we instead convert the optimization problem into an equivalent problem with spanned income but heterogeneous beliefs (see also Section 5 in \shortciteNP{CLM10}). We define the $\P$-equivalent probability measures $\P_i$, $i=1,...,I$, via the Radon-Nikodym derivative $\frac{d\P_i}{d\P} := \pi_{iT}>0$ on $\sF_T$, where
$$
\pi_{it}:= \exp\left(-\frac{\beta_{Y_i}}{\tau_i} \int_0^t \sqrt{v_u}dZ_{iu} -\frac12 \frac{\beta^2_{Y_i}}{\tau_i^2}\int_0^t v_u  du\right),\quad t\in[0,T].
$$
By Novikov's condition and Girsanov's theorem, we know that under each $\P_i$, $i=1,...,I$, the processes $W$ and $Z_{it}+ \frac{\beta_{Y_i}}{\tau_i}\int_0^t \sqrt{v_u}du$ are independent Brownian motions. We will need the processes $\tilde{Y}_{i}$ defined by $\tilde{Y}_{i0} :=  Y_{i0}$ and
\begin{align}\label{def:Y_Hspanned}
d\tilde{Y}_{it} := \mu_{Y_i} dt + \Big(\kappa_{Y_i}- \frac12 \frac{\beta^2_{Y_i}}{\tau_i}\Big) v_tdt+\sigma_{Y_i} \sqrt{v_t}dW_t,
\end{align}
for $t\in[0,T]$. By using the processes $(\pi_i,\tilde{Y}_i)$, we can re-write the objective in \eqref{eq:valuefunc} as
\begin{align*}
\E\left[\int_0^T U_i\left(c_u + Y_{iu}\right)du \right] &=
\E\left[\int_0^T \pi_{iu}U_i\left(c_u + \tilde{Y}_{iu}\right)du\right]\\
&=
\E^{\P_i}\left[\int_0^T U_i\left(c_u + \tilde{Y}_{iu}\right)du \right],
\end{align*}
where the last equality follows from the martingale property of $\pi_i$ and iterated conditional expectations. Problem \eqref{eq:valuefunc} can then be re-stated as
\begin{align}\label{valuefunc_complete}
\sup_{(c,\theta)\in \sA}
\E^{\P_i}\left[\int_0^T U_i\left(c_u + \tilde{Y}_{iu}\right)du\right],
\end{align}
which can be seen as a complete market consumption-portfolio optimization problem with the spanned income rate process $\tilde{Y}_i$ and heterogeneous beliefs $\P_i$. As detailed in the proof section, the following result follows from a variation of the martingale method for complete markets.

\begin{theorem}\label{thm:ind} Under Assumption \ref{ass:main} and under the assumption that Conjectures \ref{con:S0}, \ref{con:S} and \ref{con:S_div} hold, there exists a unique constant $\alpha_i>0$ such that
\begin{align}\label{def:lagrange}
\E\left[
%\xi_T^\text{min}\hat{X}_{i} +
\int_0^T \xi^\text{min}_u \hat{c}_{iu}du\right] = X_{i0},
\end{align}
where
\begin{align}
%\hat{X}_{i} := -\frac1{a_i}\log\left(\frac{\alpha_i\xi^\text{min}_T}{a_i}\right) - \tilde{Y}_{iT},\quad
\hat{c}_{i0} &:= -\tau_i\log\left(\tau_i\alpha_i\right) - Y_{i0},\label{optimal_c0}
\end{align}
and the consumption process has the dynamics
\begin{align}
d\hat{c}_{it} :=\Big\{\tau_i r_t + \Big(\frac12 {\tau_i\mu_S^2}+ \frac12 \frac{\beta_{Y_i}^2}{\tau_i}-\kappa_{Y_i} \Big)v_t
- \mu_{Y_i}\Big\}dt+ \Big(\tau_i\mu_S - \sigma_{Y_i}\Big)\sqrt{v_t}dW_t.\label{optimal_c}
\end{align}
Furthermore, there exists an investment strategy $\hat{\theta}_i$ such that the pair $(\hat{c}_i,\hat{\theta}_i)\in\sA$ is optimal for investor $i$, $i=1,...,I$.
\end{theorem}

\noindent The proof of this result produces the optimal investment strategy $\hat{\theta}_i$ using the martingale representation theorem (see equation \eqref{mg_rep1}) via the relation
\begin{align}\label{mg_rep}
X^{\hat{\theta}_i,\hat{c}_i}_t = \E_t^{\Q^\text{min}}\left[
%e^{-\int_t^T r_sds}\hat{X}_{i} +
\int_t^T e^{-\int_t^u r_sds}\hat{c}_{iu} du\right],\quad t\in[0,T].
\end{align}
However, a tractable expression for the optimal investment strategy $\hat{\theta}_i$ is not available because the interest rate $r_t$ is stochastic. Fortunately, our equilibrium approach only requires the abstract existence of $\hat{\theta}_i$. The proof of Theorem \ref{thm:ind} shows that the optimal strategies $(\hat{\theta}_i,\hat{c}_i)$ are such that the process \eqref{ineq:budget} is a martingale for all state-price densities $\xi^\nu$.

%From \eqref{mg_rep} we see that in the continuous case $(\Gamma_\infty)$ we have $X_{T}^{\hat{c}_i,\hat{\theta}_i}=0$, $\P$-a.s., meaning that the investors optimally leave no wealth behind after maturity $T$. On the other hand, \eqref{mg_rep} shows that in the discrete case $(\Gamma_N)$ we have $X_{T}^{\hat{c}_i,\hat{\theta}_i} = \hat{c}_{iT} \Delta$, $\P$-a.s., which means that at maturity each investor consumes whatever remaining wealth there is.

Finally, let us find the individual investors' optimal state-price densities. From the proof of Theorem \ref{thm:ind} (see equation \eqref{eq:FOC_complete}) we get the relation
$$
U_i'(\hat{c}_{iu} + Y_{iu})= \pi_{it}U_i'(\hat{c}_{iu} + \tilde{Y}_{iu}) = \alpha_i \pi_{it}\xi^\text{min}_t,
$$
where the Lagrange multiplier $\alpha_i$ is defined by \eqref{def:lagrange}. Therefore, investor $i$'s state-price density is $\hat{\xi}_{it}:=\pi_{it}\xi^\text{min}_t$, which implies that the ratios between the investors' marginals are non-constant. By the second welfare theorem, we therefore do \emph{not} expect Pareto efficiency of the equilibrium allocations, and this was indeed confirmed in Section 4.1.

\subsection{Adjusted aggregate endowment}

In order to put our equilibrium into a different perspective, let us re-consider the heterogenous beliefs formulation \eqref{valuefunc_complete}. It follows from Bayes' rule that $W$ remains a Brownian motion under each $\P_i$. Because the adjusted income processes $(\tilde{Y}_i)_{i=1}^I$ defined by  \eqref{def:Y_Hspanned} as well as the wealth dynamics $dX^{\theta,c}_t$---including $r_t$---are driven solely by $W$, we have
$$
\E^{\P_i}\left[\int_0^TU_i\left(c_u + \tilde{Y}_{iu}\right)du \right]=
\E^{\P_1}\left[\int_0^T U_i\left(c_u + \tilde{Y}_{iu}\right)du \right],
$$
for $i=1,2...,I$ and $(\theta,c)\in\sA$. Hence, if we define $\tilde{\sE}_t := \sum_{i=1}^I \tilde{Y}_{it}$ with the dynamics
$$
d\tilde{\sE}_t := \sqrt{v_t} \sigma_\sE dW_t +\big( \kappa_\sE-\frac12 \sum_{i=1}^I \frac{\beta_{Y_i}^2}{\tau_i} \big)v_t dt +\mu_\sE dt,
$$
as the economy's ``aggregate endowment'' rate, we can reduce the search for an equilibrium to a complete market equilibrium with aggregate  endowment $\tilde{\sE}_t$. In other words, by replacing \eqref{rep_prop} with the following adjusted first-order condition in the representative agent's problem
\begin{align}\label{rep_prop_adjust}
e^{-\frac1{\tau_\Sigma} \tilde{\sE}_t} \propto  \xi_t^\text{rep}, \quad t\in [0,T],
\end{align}
we recover the correct incomplete securities markets equilibrium derived in Theorem \ref{thm:main}.

\section{Model variations}
The model used for the income processes \eqref{def:v}-\eqref{def:Y_H} is chosen for its mathematical simplicity. In this section, we briefly point to a number of variations of the model, some of which are needed in order for the model to produce realistic equilibrium predictions. First of all, it is straightforward to replace the constants $(\mu_v, \kappa_v, \sigma_v, \mu_{Y_i}, \kappa_{Y_i}, \sigma_{Y_i}, \beta_{Y_i})$, $i=1,2,...,I$, describing the income dynamics with suitable deterministic functions of time. Such variations of the model are naturally required for model calibration to market data, however, the analysis is completely similar.

\subsection{Gaussian models}

We can modify our setting to produce an equilibrium in which the absolute income volatility process follows a Gaussian process. This is inspired by Stein and Stein's stochastic volatility model \citeN{SS91}, where the relative volatility process is the Gaussian process
$$
dv_t:= (\mu_v + \kappa_vv_t)dt + \sigma_v dW_t,\quad t\in[0,T],\quad v_0 >0,
$$
for $(\mu_v,\kappa_v,\sigma_v)\in\R$. Instead of the income dynamics \eqref{def:Y_H}, we consider
\begin{align*}%\label{def:Y_SS}
dY_{it} &:= (\mu_{Y_i} + \kappa_{Y_i}v_t) dt + v_t\Big(\sigma_{Y_i}dW_t + \beta_{Y_i} dZ_{it}\Big), \quad t\in[0,T],\quad Y_{i0}\in\R.
\end{align*}
In this setting, the equilibrium interest rate as well as the equilibrium instantaneous market price of risk processes will be affine functions of $v$, i.e., Gaussian processes. The resulting interest rate is Vasi\v cek's famous term structure model, whereas this type of instantaneous market price of risk process was originally used in \citeN{KO96}. Gaussian based instantaneous market price of risk models have been widely used in the finance literature (see, e.g., \citeNP{Wac02}, \citeNP{MS04} and \citeNP{BK05}).

\subsection{Terminal consumption}

Instead of running consumption, we can consider terminal consumption only. As we shall see in the next result, we need to allow $\mu_S$ in \eqref{eq:S} to be a continuous function on $[0,T]$. In this setting, the optimization problem \eqref{eq:valuefunc} is replaced by
\begin{equation}\label{eq:valuefunc_terminal}
\sup_{\theta\in \sA^\text{term}} \; \E\left[U_i(X^\theta_{iT}+Y_{iT})\right]
=\E\left[U_i(X^{\hat{\theta}_i}_{iT}+Y_{iT})\right].
\end{equation}
The wealth process $X_i^\theta$ is defined by setting $c:=0$ in \eqref{def:wealth}, i.e.,
$$
dX^\theta_{it} := r_tX^\theta_{it}dt + \theta_t\sigma_{St} \Big( \mu_S(t)\sqrt{v_t}dt + dW_t\Big),\quad X^\theta_{i0}=X_{i0}\in\R.
$$
We define the admissible strategies $\sA^\text{term}$ to be those measurable and adapted processes $\theta$ for which $X^\theta_t$ is well-defined and $X^\theta\xi^\nu$ is a supermartingale for all state-price densities $\xi^\nu$. The analogue of Theorem \ref{thm:main} is the following result.

%Because the investors only consume at time $T$ the interest rate $r$ cannot be determined in equilibrium.

\begin{theorem}\label{thm:terminal} Under Assumptions \ref{feller} and \ref{ass:main}, %, and $\sum_{i=1}^I X_{i0}=0$:
there exists an equilibrium for which Conjectures \ref{con:S0}, \ref{con:S}, and \ref{con:S_div} hold with $r_t:=0$ in \eqref{eq:S0} and $\mu_S$ in \eqref{eq:S} replaced by the deterministic function
\begin{align}\label{eq:mpr_terminal}
\mu_S(t) := \frac{\sigma_\sE}{\tau_\Sigma}-b(T-t)\sigma_v,\quad t\in [0,T],
\end{align}
where $b$ is defined by the Riccati equation \eqref{ode:b}.
\end{theorem}

\noindent In this setting of consumption at time $T$ only, the interest rate cannot be determined in equilibrium, and we choose $r_t:=0$ for simplicity. Consequently, the minimum martingale measure $\Q^\text{min}$ and the minimum forward measure $\Q^U$ are identical, and the instantaneous market price of risk process is identical to the discrete market price of risk process measured over finite time-intervals $[0,U]$, $U\in[0,T]$.

Contrary to Theorem \ref{thm:cont}, the setting of terminal consumption only  produces an impact on the instantaneous market price of risk process due to income incompleteness. In order to see this, we proceed as in Section 4.1 except that the first-order-condition \eqref{rep_prop} is only required to hold at $t=T$. To compute the instantaneous market price of risk process corresponding to the representative agent, we need the dynamics of the martingale $\xi_t^\text{rep} := \E_t[e^{-\frac1{\tau_\Sigma}\sE_T}]/\E[e^{-\frac1{\tau_\Sigma}\sE_T}]$ for $t\in[0,T]$. Similarly to the proof of Lemma \ref{lem:ZCB} we have
\begin{align}\label{eq:rep_agent_terminal}
d\xi_t^\text{rep} = -\xi_t^\text{rep}\sqrt{v_t} \Big(\Big\{\frac{\sigma_\sE}{\tau_\Sigma}-b_\text{rep}(T-t)\sigma_v\Big\} dW_t + \frac1{\tau_\Sigma}\sum_{i=1}^I \beta_{Y_i}dZ_{it}\Big),
\end{align}
where $b_\text{rep}$ is defined in Section 4.1. By comparing \eqref{eq:mpr_terminal} and the $dW$-coefficient in \eqref{eq:rep_agent_terminal} we see from Section 4.1 that the incompleteness impact on the instantaneous market price of risk process in the case of only terminal consumption at time $T$ is identical to the impact on the discrete market price of risk process measured over the interval $[0,T]$ in the case of continuous consumption.

Because the equilibrium interest rate is zero, the investors value functions will be of the exponential-affine form. Consequently, the investors' optimal trading strategies $\hat{\theta}_{it}$ can be computed explicitly using HJB-techniques.

In conclusion, this section shows that it is not the difference between the minimal martingale and forward measures due to equilibrium stochastic interest rates which produces the impact of unspanned income on the discrete market price of risk process. Instead, the key observation is that the impact of stochastic unspanned income volatility must be integrated over a time interval in order to produce an effect on the market price of risk process such as returns measured over finite time-intervals (Section 4.1) or consumption only taking place at discrete points in time.

\section*{Appendix: Proofs}
\renewcommand{\thesection}{\Alph{section}}
\setcounter{section}{1}

We start by proving Lemma \ref{lem:ZCB} since this result is used in the later proofs. Then we state and prove a result which is used in the proof of Theorem \ref{thm:ind} for individual optimality, given Conjectures \ref{con:S0}, \ref{con:S}, and \ref{con:S_div} about the spot interest rate and the risky security price processes. We then prove Theorem \ref{thm:main} stating that the conjectured spot interest rate and the risky security price processes indeed clear all markets. The proof of Theorem \ref{thm:terminal} is similar to the proof of Theorem \ref{thm:main} and we will only outline the few  differences. Finally, we prove Theorem \ref{thm:cont} stating that there is no impact of incompleteness on the instantaneous market price of risk process in settings based on exponential investors and continuous income processes governed by Brownian motions.

\proof[Proof of Lemma \ref{lem:ZCB}]  The discriminant corresponding to the Riccati equation \eqref{ode:b} is defined as
$$
q := (\kappa_v-\frac{\sigma_\sE}{\tau_\Sigma}\sigma_v)^2 -2\sigma_v^2\frac1{\tau_\Sigma} \Big( \sum_{i=1}^I \frac{\beta_{Y_i}^2}{2\tau_i} + \frac{\sigma_\sE^2}{2\tau_\Sigma} -\kappa_\sE  \Big).
$$
Under Assumption \ref{ass:main}, $q$ is positive. The appendix in \citeN{KO96} on normal Riccati equations ensures that \eqref{ode:b} has a non-exploding unique normal solution $b$ with $b(s)\neq 0$ for $s\in(0,\infty)$.

In order to calculate the zero-coupon bond prices, we need the dynamics of the volatility process $v$ defined by \eqref{def:v} under the minimal martingale measure $\Q^\text{min}$:
\begin{align}\label{dv_under_Qmin}
dv_t = \Big(\mu_v +(\kappa_v-\mu_S\sigma_v)v_t\Big)dt + \sigma_v\sqrt{v_t} dW^{\Q^\text{min}}_t.
\end{align}
Therefore, $v_t$ is also a Feller process under the minimal measure $\Q^\text{min}$. By It\^o's lemma we see that the process $N_t := \exp\big( b(U-t) v_t-a(U-t) \big)/S_t^{(0)}$ is a local martingale under $\Q^\text{min}$ which has the dynamics
$$
dN_t = N_t b(U-t) \sigma_v \sqrt{v_t} dW^\text{min}_t, \quad N_U=1/S_U^{(0)}.
$$
Since $v_t$ has a non-central $\chi^2$-distribution and $b$ is a bounded continuous function, Novikov's condition is satisfied locally on $[0,U]$. We can then use the argument on p.233 in \citeN{LS77} to see that $N$ is a martingale on $[0,U]$. This martingale property and the terminal condition $N_U=1/S_U^{(0)}$ show that $B(t,U) = N_t$ for $t\in[0,U]$ and the claim follows.

$\endproof$

In the later proofs we will need the following result, where the main complication is that $\nu$ can depend on both $W$ and $W^\perp$ and, hence, the random variable $\int_0^T \nu_u dW^\perp_u$ is \emph{not} independent of $\sF^W_t:= \sigma(W_u)_{u\in[0,t]}$.

\begin{lemma}\label{lem:independence} Under the assumptions of Theorem \ref{thm:ind}: Let $\xi^\nu_t = M_t^\nu/S_t^{(0)}$ be a state-price density as in Definition \ref{def:state_price}. Then for $0\le s\le t\le T$ we have
$$
\E[\,M^\nu_t \,|\,\sF_s \vee \sF_t^W] = M_s^\nu \frac{M^\text{min}_t}{M^\text{min}_s} ,\quad \P\text{-a.s.},
$$
where $\xi^\text{min}_t = M_t^\text{min}/S_t^{(0)}$ is the minimal state-price density.
\end{lemma}

\proof By the definition of a state-price density $\xi^\nu$, we can find a $W$-independent Brownian motion $W^\perp$ as well as $\nu\in\sL^2$ such that
$$
dM^\nu_u = - M^\nu_u \Big( \mu_S \sqrt{v_u} dW_u + \nu_u dW^\perp_u\Big), \quad M_0^\nu =1,
$$
is a martingale. We define the corresponding $\P$-equivalent probability measure $\Q^\nu$  by $\frac{d\Q^\nu}{d\P} := M_T^\nu$ on $\sF_T$. $\Q^{\min}$ denotes the minimal martingale measure under which $W_v^{\Q^\text{min}}:= W_v + \mu_S\int_0^v\sqrt{v_u}du$ is a Brownian motion. By Girsanov's theorem,  $W^{\Q^\text{min}}$ is also a $\Q^\nu$-Brownian motion.

For a set $A_t\in \sF_t^W$ the martingale representation theorem produces an $\sF^W$-adapted process $f\in\sL^2$ such that
$$
g_v := \E^{\Q^{\text{min}}}[1_{A_t}\,|\,\sF_v] = \E^{\Q^{\text{min}}}[1_{A_t}] + \int_0^v f_u dW^{\Q^\text{min}}_u,\quad v\in[0,t].
$$
Since $1_{A_t}$ is bounded, the process $g_v$ is a bounded $\Q^{\text{min}}$-martingale. Furthermore, since $W^{\Q^\text{min}}$ is also a $\Q^\nu$-Brownian motion, $g_v$ is a local $\Q^\nu$-martingale. However, by $g_v$'s boundedness property, $g_v$ is actually a $\Q^\nu$-martingale.

To conclude the proof, we let $A_s\in\sF_s$ be arbitrary. Then we have
\begin{align*}
\E[M^\nu_t 1_{A_t}1_{A_s}] &= \E[1_{A_s}\,\E[M^\nu_t g_t\,|\,\sF_s]\,] \\&= \E[1_{A_s}\,\E^{\Q^\nu}[ \,g_t\,|\,\sF_s]\,M^\nu_s\,]\\&= \E[1_{A_s}g_sM^\nu_s\,] \\&= \E\left[1_{A_s}\frac{\E[M_t^\text{min}1_{A_t}\,|\, \sF_s]}{M_s^\text{min}}\,M^\nu_s\,\right]= \E\left[1_{A_s}1_{A_t}\frac{M_t^\text{min} }{M_s^\text{min}}\,M^\nu_s\,\right].
\end{align*}
The first equality follows from iterated expectations and the $\sF_s$-measurability of $A_s$. The second equality is Bayes' rule for conditional expectations. The third equality is $g$'s martingale property under $\Q^\nu$. The fourth equality is again Bayes' rule, whereas the last equality is produced by the $\sF_s$-measurability of $A_s, M^\nu_s, M_s^\text{min}$ and iterated expectations. The arbitrariness of $A_t\in\sF_t^W,A_s\in\sF_s$ and the $\sF_s\vee \sF_t^W$-measurability of $\frac{M_t^\text{min} }{M_s^\text{min}}\,M^\nu_s$ conclude the proof.

$\endproof$

\noindent \proof[Proof of Theorem \ref{thm:ind}] %Let us first consider the complete market problem of maximizing \eqref{valuefunc_complete} over measurable and adapted processes $(\theta,c)$ satisfying \eqref{terminal_pos} and producing supermartingality under the subjective probability measure $\P_i$ of the process
%\begin{align}\label{ineq:complete_budget}
%\E^{\Q^\text{min}}\left[\int_0^T\exp\left(-\int_0^u r_sds\right) c_{iu}\Gamma(du)\right] =
%\xi^\text{min}_t X^{\theta,c}_t + \int_0^t\xi^\text{min}_u c_{u}du,\quad t\in[0,T].
%\end{align}

Let $\hat{c}_i$ be defined as in the theorem's statement. Our first task is to show that the following process is well-defined (we note that $\hat{X}_{iT}=0$)
\begin{align}\label{mg_rep1}
\hat{X}_{it} := \E_t^{\Q^\text{min}}\left[
%e^{-\int_t^T r_sds}\hat{X}_{i}+
\int_t^T e^{-\int_t^u r_sds} \hat{c}_{iu} du\right],\quad t\in[0,T].
\end{align}
Under Assumption \ref{ass:main}, we can find a constant $p>1$ such that
\begin{align}\label{ass:r_neg_restriction1}
(\kappa_v -\frac{\sigma_\sE}{\tau_\Sigma}\sigma_v)^2 > 2p\sigma_v^2\frac1{\tau_\Sigma}\Big( \sum_{i=1}^I \frac{\beta_{Y_i}^2}{2\tau_i} + \frac{\sigma_\sE^2}{2\tau_\Sigma} -\kappa_\sE\Big).
\end{align}
We then consider the coupled system of ODEs for $s\in(0,\infty)$
\begin{align*}
\tilde{b}'(s) &= \tilde{b}(s)(\kappa_v-\frac{\sigma_\sE}{\tau_\Sigma}\sigma_v) +\frac12\tilde{b}(s)^2\sigma_v^2 +\frac{p}{\tau_\Sigma} \Big( \sum_{i=1}^I \frac{\beta_{Y_i}^2}{2\tau_i} + \frac{\sigma_\sE^2}{2\tau_\Sigma} -\kappa_\sE  \Big),\quad \tilde{b}(0)=0,\\
\tilde{a}'(s) &= p\frac{\mu_\sE}{\tau_\Sigma}-\tilde{b}(s)\mu_v,\quad \tilde{a}(0)=0.
\end{align*}
The restriction \eqref{ass:r_neg_restriction1} ensures a positive discriminant corresponding to $\tilde{b}$'s ODE. Therefore, the appendix in \citeN{KO96} on normal solutions ensures that $\tilde{b}$ and, hence, also $\tilde{a}$, is a continuous function on $[0,\infty)$. Arguing as in the proof of Lemma \ref{lem:ZCB}, we find
$$
\E_t^{\Q^\text{min}}\left[e^{-p\int_t^u r_sds}\right] = e^{\tilde{b}(u-t)v_t-\tilde{a}(u-t)},\quad 0\le t\le u\le T.
$$
To verify that \eqref{mg_rep1} indeed is well-defined, we can use Tonelli's theorem to write
\begin{align*}
\E^{\Q^\text{min}}\left[
\int_t^T e^{-\int_t^u r_sds} |\hat{c}_{iu}| du\right] = \int_t^T\E^{\Q^\text{min}}\left[e^{-\int_t^u r_sds}  |\hat{c}_{iu}| \right]du.
\end{align*}
By H\"older's inequality it therefore suffices to show that the expectations
\begin{align}\label{CS_estimate}
\E^{\Q^\text{min}}\left[e^{-p\int_t^u r_sds}\right]\quad \text{and}\quad \E^{\Q^\text{min}}\left[|\hat{c}_{iu}|^{\frac{p}{p-1}}\right],
\end{align}
are bounded uniformly in $u\in[t,T]$. We start with the first term. If \eqref{ass:r_neg} holds, $r_t$ defined by \eqref{r} is bounded from below, and the claim follows since $p>1$. On the other hand, if \eqref{ass:r_neg} fails, we get the inequality for $u\in[t,T]$
\begin{align*}
e^{\tilde{b}(u)v_0-\tilde{a}(u)}&=\E^{\Q^\text{min}}\left[e^{-p\int_0^u r_sds}\right] \\&=  \E^{\Q^\text{min}}\left[e^{-p\frac{\mu_\sE}{\tau_\Sigma}u+\frac{p}{\tau_\Sigma}\left( \sum_{i=1}^I \frac{\beta_{Y_i}^2}{2\tau_i} +\frac{\sigma_\sE^2}{2\tau_\Sigma} -\kappa_\sE\right)\int_0^u v_sds}\right]
\\&\ge  e^{-p\frac{\mu_\sE}{\tau_\Sigma}t} \;\E^{\Q^\text{min}}\left[e^{-p\frac{\mu_\sE}{\tau_\Sigma}(u-t)+\frac{p}{\tau_\Sigma}\left( \sum_{i=1}^I \frac{ \beta_{Y_i}^2}{2\tau_i} +\frac{\sigma_\sE^2}{2\tau_\Sigma} -\kappa_\sE\right)\int_t^u v_sds}\right] \\&=e^{-p\frac{\mu_\sE}{\tau_\Sigma}t} \;\E^{\Q^\text{min}}\left[e^{-p\int_t^u r_sds}\right].
\end{align*}
Since both $\tilde{a}$ and $\tilde{b}$ are continuous functions on $[0,T]$ and, hence, bounded, we obtain
$$
 \E^{\Q^\text{min}}\left[e^{-p\int_t^u r_sds}\right]\le e^{p\frac{\mu_\sE}{\tau_\Sigma}t}\;\max_{s\in[0,T]} e^{\tilde{b}(s)v_0-\tilde{a}(s)}<\infty,\quad u\in[t,T].
$$

We will next provide a uniform bound (in $u\in[0,T]$) of the second term in \eqref{CS_estimate}. In the following argument $C_1,C_2,...$ denote various irrelevant positive constants. Since $v_t>0$, we have the following chain of inequalities
\begin{align*}
\E^{\Q^\text{min}}&\left[|\hat{c}_{iu}|^{\frac{p}{p-1}}\right] \\& \le C_1 + C_2\E^{\Q^\text{min}}\left[\left(\int_0^u v_sds\right)^{\frac{p}{p-1}}\right] + C_3\E^{\Q^\text{min}}\left[\left|\int_0^u \sqrt{v_s}dW^{\Q^\text{min}}_s\right|^{\frac{p}{p-1}}\right]\\
& \le C_1 + C_2u^{\frac1{p-1}}\int_0^u \E^{\Q^\text{min}}\left[v_s^{\frac{p}{p-1}}\right]ds + C_4\E^{\Q^\text{min}}\left[\left(\int_0^u v_sds\right)^{\frac12\frac{p}{(p-1)}}\right]\\
& \le C_1 + C_2T^{\frac1{p-1}}\int_0^T \E^{\Q^\text{min}}\left[v_s^{\frac{p}{p-1}}\right]ds + C_4\E^{\Q^\text{min}}\left[\left(\int_0^u v_sds\right)^{\frac{p}{(p-1)}}\right]^\frac12\\
& \le C_1 + C_2T^{\frac1{p-1}}\int_0^T \E^{\Q^\text{min}}\left[v_s^{\frac{p}{p-1}}\right]ds + C_4\left(T^{\frac1{p-1}}\int_0^T \E^{\Q^\text{min}}\left[v_s^{\frac{p}{p-1}}\right]ds\right)^\frac12.
\end{align*}
The first inequality follows from the definition of $\hat{c}_i$. The second inequality uses Jensen's inequality (recall $p>1$) and Tonelli's theorem on the $ds$-integral, whereas the estimate of the $dW^{\Q^{\text{min}}}$-integral follows from the Burkholder-Davis-Gundy inequality (see, e.g., Theorem 3.28 on p.166 in \citeN{KS88}).  The third inequality first uses $u\le T$ and Jensen's inequality on the second $ds$-integral. The last estimate is similar. The dynamics \eqref{dv_under_Qmin} for $v$ ensure that $v_s$ is non-central $\chi^2$-distributed under $\Q^\text{min}$ and, hence, the $ds$-integrals are finite. All in all, we have shown that when Assumption \eqref{ass:main} holds, the process \eqref{mg_rep1} is well-defined and finite.

We next establish the existence of $\alpha_i>0$ satisfying \eqref{def:lagrange}. The requirement \eqref{def:lagrange} becomes
\begin{align}\label{def:lagrange1}
X_{i0} = S_0\hat{c}_{i0} + \E^{\Q^\text{min}}\left[
\int_0^T e^{-\int_0^u r_sds}\left(\int_0^u d\hat{c}_{is}\right) du\right].
\end{align}
Inserting the definition of $\hat{c}_{i0}$ from \eqref{optimal_c0} into \eqref{def:lagrange1} produces an equation uniquely characterizing the Lagrange multiplier $\alpha_i\in(0,\infty)$.

We now turn to the admissibility requirement. Using the relation between $\Q^\text{min}$ and $\xi^\text{min}$ and $\hat{X}_{it}$'s definition produce for $t\in[0,T]$
\begin{align*}
\frac{\hat{X}_{it}}{S_t^{(0)}} + \int_0^t\frac{\hat{c}_{iu}}{S_u^{(0)}} du &= \E_t^{\Q^\text{min}}\left[\int_0^T \frac{\hat{c}_{iu}}{S_u^{(0)}} du\right]\\& = \E^{\Q^\text{min}}\left[\int_0^T \frac{\hat{c}_{iu}}{S_u^{(0)}} du\right] + \int_0^t f_{iu} dW^{\Q^\text{min}}_u\\
&= X_{i0}+\int_0^t \frac1{S_u^{(0)}}\hat{\theta}_{iu}\sigma_{Su}dW^{\Q^\text{min}}_u.
\end{align*}
The integrand $f_i\in\sL^2$ appearing in the second equality comes from the martingale representation theorem for $\sF^W_t:=\sigma(W_u)_{u\in[0,t]}$ after noticing that all involved quantities are $\sF^W_t$-adapted. The last equality follows from \eqref{def:lagrange} and by defining
$$
\hat{\theta}_{iu} := \frac{S^{(0)}_uf_{iu}}{\sigma_{Su}},\quad u\in[0,T),
$$
which is possible since we are assuming that Conjecture \ref{con:S} holds. All in all, this shows that $\hat{X}_{it}$ has the form $\eqref{def:wealth}$ and that
\begin{align}\label{ineq:complete_budget}
%\E^{\Q^\text{min}}\left[\int_0^T\exp\left(-\int_0^u r_sds\right) c_{iu}\Gamma(du)\right] =
\xi^\text{min}_t \hat{X}_{it} + \int_0^t\xi^\text{min}_u \hat{c}_{iu}du,\quad t\in[0,T].
\end{align}
is a $\P$-martingale. We will next show the supermartingale requirement  \eqref{ineq:budget} by proving the stronger martingale property.
%We finally prove the optimality of $(\hat{\theta}_i,\hat{c}_i)$ for problem \eqref{eq:valuefunc} and problem \eqref{valuefunc_complete}. By noticing that $\pi_{it} \xi^\text{min}_t$ is a state-price density, we see that the $\P_i$-supermartingale requirement \eqref{ineq:complete_budget} is weaker than the $\P$-supermartingale requirement \eqref{ineq:budget}. Therefore,  the proof is completed by verifying that the pair $(\hat{\theta}_i,\hat{c}_i)$ also satisfies \eqref{ineq:budget}.
By the definition of a state-price density $\xi^\nu_t$, we can find a martingale $M^\nu$ such that $\xi^\nu_t = M_t^\nu/S_t^{(0)}$. We then get for $0\le s \le t\le T$
\begin{align*}%\label{property:supermg}
\E\left[\xi^\nu_t\hat{X}_{it} + \int_0^t \xi_u^\nu\hat{c}_{iu}\,du\Big|\sF_s\right]&=\E\left[M_t^\nu\Big(\hat{X}_{it}/S_t^{(0)} + \int_0^t \hat{c}_{iu}/S_u^{(0)}\,du\Big)\Big|\sF_s\right]\\&=\E\left[\E\left[M_t^\nu \Big| \sF_s\vee \sF^W_t\right]\Big(\hat{X}_{it}/S_t^{(0)} + \int_0^t \hat{c}_{iu}/S_u^{(0)}\,du\Big)\Big|\sF_s\right]
\\&=\frac{M_s^\nu}{M_s^\text{min}}\E\left[M_t^\text{min}\Big(\hat{X}_{it}/S_t^{(0)} + \int_0^t \hat{c}_{iu}/S_u^{(0)}\,du\Big)\Big|\sF_s\right] \\&=\xi^\nu_s\hat{X}_{is} + \int_0^s \xi_u^\nu\hat{c}_{iu}\,du.
\end{align*}
The first equality follows from $\xi^\nu = M^\nu/S^{(0)}$ and the martingale property of $M^\nu$. The second equality follows from iterated expectations and the $\sF^W_t$-measurability of $\hat{X}_{it},(\hat{c}_{iu})_{u\in[0,t]}$ and $(S^{(0)}_u)_{u\in[0,t]}$. The third equality uses Lemma \ref{lem:independence}, whereas the last equality is $\xi_t^\text{min} = M_t^\text{min}/S_t^{(0)}$ combined with the already established martingale property of \eqref{ineq:complete_budget}. This shows $(\hat{\theta}_i,\hat{c}_i)\in\sA$.

Finally, we will verify the optimality of $(\hat{\theta}_i,\hat{c}_i)$ for problem \eqref{valuefunc_complete} and, hence, also for problem \eqref{eq:valuefunc}. For the case of positive wealth processes the standard argument can be found in Section 3.6 in \citeN{KS98}. In the following, $V$ denotes the convex conjugate of $U$ (see Section 3.4 in \citeNP{KS98}). By Fenchel's inequality, we have
\begin{align*}
&U(c_u +  \tilde{Y}_{iu}) \le V(\alpha_i\, \xi^\text{min}_u) + \alpha_i\, \xi^\text{min}_u\Big(c_u +  \tilde{Y}_{iu}\Big), \quad u\in[0,T].
%&U(X^{c,\theta}_{iT} +  \tilde{Y}_{iT}) \le V(\alpha_i\, \xi^\text{min}_T) + \alpha_i\, \xi^\text{min}_T\Big(X^{c,\theta}_{iT} +  \tilde{Y}_{iT}\Big).
\end{align*}
Integrating with respect to $du$ and adding the positive random variable $\alpha_i\, \xi^\text{min}_TX^{c,\theta}_{iT}$, see \eqref{terminal_pos}, give us
\begin{align*}
\int_0^TU(c_u +  \tilde{Y}_{iu}) du\le  \int_0^T\Big\{V(\alpha_i
 \xi^\text{min}_u) + \alpha_i\, \xi^\text{min}_u\Big(c_u +  \tilde{Y}_{iu}\Big)\Big\}du  +\alpha_i\, \xi^\text{min}_TX^{c,\theta}_{iT}.
\end{align*}
Since $\pi_{it}\xi^\text{min}_t$ is a state-price density and the subjective probability measure $\P_i$ is defined by $\frac{d\P_i}{d\P} := \pi_{iT}$, we can use the supermartingale property \eqref{ineq:budget} to obtain the inequality
\begin{align*}
\E^{\P_i}\left[\int_0^TU(c_u +  \tilde{Y}_{iu}) du\right]
&\le \E^{\P_i}\left[\int_0^T\Big\{V(\alpha_i\, \xi^\text{min}_u) + \alpha_i\, \xi^\text{min}_u\tilde{Y}_{iu}\Big\}du\right] + \alpha_i X_{i0}\\&=\E^{\P_i}\left[\int_0^T\Big\{V(\alpha_i\, \xi^\text{min}_u) + \alpha_i\, \xi^\text{min}_u\tilde{Y}_{iu}\Big\}du\right] + \alpha_i \E^{\P_i}\left[\int_0^T\xi^\text{min}_u \hat{c}_{iu}du\right]\\
&=
\E^{\P_i}\left[\int_0^TU(\hat{c}_{iu} +  \tilde{Y}_{iu}) du\right].
\end{align*}
The first equality follows from the established martingale property, whereas the last equality follows from the first-order condition
\begin{align}\label{eq:FOC_complete}
U_i'(\hat{c}_{iu} + \tilde{Y}_{iu}) = \alpha_i\, \xi^\text{min}_u, \quad u\in[0,T],
\end{align}
and the relation between $U$ and $V$ stated in Lemma 4.3(i) in \citeN{KS98}. In order to verify that \eqref{eq:FOC_complete} holds, we use \eqref{optimal_c0} to see that \eqref{eq:FOC_complete} holds for $u=0$. Furthermore, by using \eqref{optimal_c} we see that the dynamics of both sides of \eqref{eq:FOC_complete} are identical and, hence, \eqref{eq:FOC_complete} holds for all $u\in[0,T]$.

$\endproof$

\noindent \emph{Proof of Theorem \ref{thm:main}.}
We define $S^{(0)}$ by \eqref{eq:S0} and $S$ by \eqref{def:S}. The already proven Lemma \ref{lem:ZCB} produces the zero-coupon bond dynamics
$$
dB(t,U) = B(t,U)\Big(r_tdt + b(U-t)\sigma_v\sqrt{v_t} dW^{\Q^\text{min}}_t\Big).
$$
We can then use Tonelli's theorem to re-write \eqref{def:S} as follows
\begin{align*}
S_t = \int_t^T \E_t^{\Q^\text{min}}[e^{-\int_t^U r_sds}]dU = \int_t^T B(t,U)dU,\quad t\in[0,T].
\end{align*}
Leibnitz' rule for stochastic integrals produces the dynamics
\begin{align*}
dS_t &= -B(t,t)dt +r_t\int_t^T B(t,U)dU dt +  \sigma_v \sqrt{v_t} \int_t^T B(t,U)b(U-t)dU\,dW^{\Q^\text{min}}_t\\
 &= -dt + r_t S_tdt + \sigma_v \sqrt{v_t} \int_t^T B(t,U)b(U-t)dU\,dW^{\Q^\text{min}}_t.
\end{align*}
Therefore, Conjecture \ref{con:S} holds with the volatility coefficient \eqref{eq:vol_coefficient}.

We now establish clearing in the goods market.  By summing up the expressions for $d\hat{c}_{it}$, we find $d\sum_{i=1}^I\hat{c}_{it} =0$, see \eqref{r}-\eqref{mu_S} for the definitions of $r_t$ and $\mu_S$. Since $\sum_{i=1}^IX_{i0}=0$, we see from \eqref{def:lagrange1} that $\sum_{i=1}^I \hat{c}_{i0} = 0$ and, hence, the goods market clears.

To see that the risky security market also clears, we sum over $i=1,...,I$ in \eqref{mg_rep} to see $\sum_{i=1}^IX^{\hat{\theta}_i,\hat{c}_i}_t=0$. By dividing both sides in this relation by $S^{(0)}$, we find the $\Q^\text{min}$-dynamics
\begin{align*}
%\frac1{S^{(0)}_t}\left( \sigma_{St} dW^{\Q^\text{min}}_t - dt\right) &= d\frac{S_t}{S^{(0)}_t}\\
0=d\sum_{i=1}^I\frac{X^{\hat{\theta}_i,\hat{c}_i}_t}{S^{(0)}_t}&= \sum_{i=1}^I\frac1{S^{(0)}_t}\Big( \hat{\theta}_{it}\sigma_{St} dW^{\Q^\text{min}}_t - \hat{c}_{it}dt\Big)
\\&= \frac1{S^{(0)}_t}\Big(\sum_{i=1}^I\hat{\theta}_{it}\Big) \sigma_{St} dW^{\Q^\text{min}}_t.
\end{align*}
The second equality follows from the definition of $W^{\Q^\text{min}}$, and  the last equality is due to clearing in the goods market. By matching the $dW^{\Q^\text{min}}$-coefficients and using $\frac1{S^{(0)}}\sigma_S\neq0$, we obtain the clearing condition.

Finally, to show clearing in the money market, we use
$$
0=\sum_{i=1}^IX^{\hat{\theta}_i,\hat{c}_i}_t=\sum_{i=1}^I\Big(\hat{\theta}_{it}S_t +\hat{\theta}^{(0)}_{it}S^{(0)}_t\Big) = S^{(0)}_t\sum_{i=1}^I\hat{\theta}^{(0)}_{it}.
$$
The first equality was established above, whereas the last equality follows from the already established clearing in the risky security market. Since $S^{(0)}>0$, the clearing condition in the money market follows.

$\endproof$

\noindent \emph{Proof of Theorem \ref{thm:terminal}.} In this setting the minimal martingale measure $\Q^{\text{min}}$ on $\sF_T$ is defined by $\frac{d\Q^{\text{min}}}{d\P}:= \frac{\xi_T^\text{min}}{\xi_0^\text{min}}$, where $\xi_t^\text{min}>0$ is the martingale
$$
d\xi_t^\text{min}:= -\xi_t^\text{min} \mu_S(t)\sqrt{v_t} dW_t,\quad t\in[0,T],\quad \xi_0^\text{min}:=1.
$$
Let $\tilde{Y}_i$ be defined by \eqref{def:Y_Hspanned}. Similarly to the proof of Theorem \ref{thm:ind}, we can use the martingale representation theorem to produce $\hat{\theta}_i\in\sA^\text{term}$ such that the corresponding wealth process $(X_t^{\hat{\theta}_i})_{t\in[0,T]}$ satisfies the first-order condition
$$
U_i'(X_T^{\hat{\theta}_i}+\tilde{Y}_{iT}) = \alpha_i \xi_T^\text{min}.
$$
Here $\alpha_i$ is the Lagrange multiplier corresponding to the budget constraint, i.e., $\alpha_i>0$ satisfies the analogue of \eqref{def:lagrange1}:
\begin{align}\label{def:lagrange2}
X_{i0} = \E^{\Q^\text{min}}\left[X_T^{\hat{\theta}_i}\right]= \E^{\Q^\text{min}}\left[-\tau_i\log(\tau_i\alpha_i \xi^\text{min}_T)-\tilde{Y}_{iT}\right].
\end{align}

In order to see that all markets clear we introduce the martingale for $t\in[0,T]$
\begin{align*}
N_t:=\tau_\Sigma\sigma_v\int_0^tb(T-u) \sqrt{v_u} dW_u - \E_t\Big[\int_0^T\Big(\kappa_{\sE} -\sum_{i=1}^I\frac{\beta_{Y_i}^2}{2\tau_i}-\frac{\tau_\Sigma\mu_S(u)^2}2  \Big)v_u du\Big].
\end{align*}
By using Fubini's theorem for conditional expectations we find the dynamics
\begin{align*}
dN_t&=\sigma_v\Big(\tau_\Sigma b(T-t)  -\int_t^T\Big\{\kappa_{\sE} -\sum_{i=1}^I\frac{\beta_{Y_i}^2}{2\tau_i}-\frac{\tau_\Sigma}2 \mu_S(u)^2 \Big\} e^{-\kappa_v(t-u)}du\Big)\sqrt{v_t}dW_t\\
&=\sigma_v\tau_\Sigma\Big( b(T-t)  +\int_t^T\Big\{b'(T-u)-b(T-u)\kappa_v \Big\} e^{-\kappa_v(t-u)}du\Big)\sqrt{v_t}dW_t,
\end{align*}
where the second equality follows from \eqref{ode:b}. However, by using integration by parts together with $b(0)=0$ we obtain $dN_t=0$ for $t\in[0,T]$.

We can then finish the proof and as in the proof of Theorem \ref{thm:main} it suffices to show $\sum_{i=1}^I X_T^{\hat{\theta}_i}=0$, $\P$-a.s., to ensure clearing in all markets. By the definition of $\xi_t^\text{min}$ the requirement $\sum_{i=1}^I X_T^{\hat{\theta}_i}=0$, $\P$-a.s., is equivalent to
\begin{align*}
\sum_{i=1}^I \tilde{Y}_{iT} &= -\sum_{i=1}^I\tau_i\log(\alpha_i\tau_i \xi_T^\text{min})\\
&=-\sum_{i=1}^I\tau_i\log(\alpha_i\tau_i)+\tau_\Sigma\left(\int_0^T \mu_S(t) \sqrt{v_t} dW_t +\frac12\int_0^T \mu_S(t)^2 v_t dt\right).
\end{align*}
By the definitions of $\mu_S$ and $\tilde{Y}_i$ this requirement is equivalent to
\begin{align*}
\sum_{i=1}^I& \left(\tilde{Y}_{i0} + \mu_{Y_i}T + \int_0^T(\kappa_{Y_i} -\frac{\beta_{Y_i}^2}{2\tau_i})v_t dt\right) \\
&=-\sum_{i=1}^I\tau_i\log(\alpha_i\tau_i)+\tau_\Sigma\left(\sigma_v\int_0^Tb(T-t) \sqrt{v_t} dW_t +\frac12\int_0^T \mu_S(t)^2 v_t dt\right).
\end{align*}
By the definition of the martingale $N$, this requirement can be re-written as
\begin{align}\label{xi0_requirement_terminal}
\sum_{i=1}^I\tilde{Y}_{i0} + \mu_{\sE}T+\sum_{i=1}^I\tau_i\log(\alpha_i\tau_i)=N_T = N_0.
\end{align}
Since $\sum_{i=1}^I X_{i0}=0$, the requirement \eqref{xi0_requirement_terminal} holds by \eqref{def:lagrange2}.

$\endproof$

\ \\
\noindent \emph{Proof of Theorem \ref{thm:cont}.} In the first-order condition for the individual investor \eqref{ind:FOC}, the investor-specific state-price density $\hat{\xi}_{i}$ has the dynamics
$$
d\hat{\xi}_{it} = -\hat{\xi}_{it}\Big( r_t dt + \lambda'_t dB_t + dM_{it}^\perp\Big),\quad i=1,...,I,
$$
for some local martingale $M_i^\perp$ orthogonal to $B$, i.e., $\langle B,M_i^\perp\rangle_t=0$ for all $t\in[0,T]$. Computing the dynamics of both sides of \eqref{ind:FOC} gives us the relation
$$
d\hat{c_i}_t = ... dt + \left(\tau_i \lambda'_t - \sigma_{Y_i't}\right) dB_t + ... dB^\perp_{it}+...dM^\perp_{it}.
$$
By summing over investors and matching the $dB$-integrals, we see that the equilibrium instantaneous market price of risk process satisfies
\begin{align*}
\lambda'_t = \frac1{\tau_\Sigma} \sum_{i=1}^I\sigma_{Y_i't},\quad t\in[0,T].
\end{align*}
$\endproof$

\bibliographystyle{chicago}
\bibliography{CLbib}

\begin{comment}

\end{comment}
\end{document}